\documentclass[preprints,review,submit,pdftex]{Definitions/mdpi} 
\firstpage{1} 
\pubvolume{1}
\issuenum{1}
\articlenumber{0}
\pubyear{2022}
\datepublished{} 
\hreflink{https://doi.org/} % If needed use \linebreak
\pdfoutput=1

\Title{{Status and Perspectives of Continuous Gravitational \mbox{Wave Searches}}} 
\Author{Ornella Juliana Piccinni $^{1,2}$\orcidA{}} %MDPI: Please carefully check the accuracy of names and affiliations.

\address{%
$^{1}$ \quad INFN, Sezione di Roma, I-00185 Roma, Italy; ornella.juliana.piccinni@roma1.infn.it %\textcolor{red}{(These two mail are inconsistent with the system, please confirm whether they need to be reserved and whether they are correct)}
\\ 
$^{2}$ \quad Institut de F\'{\i}sica d'Altes Energies (IFAE), Barcelona, Institute of Science and Technology and ICREA, E-08193 Barcelona, Spain; opiccinni@ifae.es  % \textcolor{red}{(1. Two Institut please confirm should be divided into two Affiliations 2. if and or comma should remove)}
}

\abstract{The birth of gravitational wave astronomy was triggered by the first detection of a signal produced by the merger of two compact objects (also known as a compact binary coalescence event). The following detections made by the Earth-based network of advanced interferometers had a significant impact in many fields of science: astrophysics, cosmology, nuclear physics and fundamental physics. However, compact binary coalescence signals are not the only type of gravitational waves potentially detectable by LIGO, Virgo, and KAGRA. An interesting family of still undetected signals, and the ones that are considered in this review, are the so-called continuous waves, paradigmatically exemplified by the gravitational radiation emitted by galactic, fast-spinning isolated neutron stars with a certain degree of asymmetry in their mass distribution. In this work, I will review the status and the latest results from the analyses of advanced detector data, including searches for dark matter signatures.}

\begin{document}
%\begin{paracol}
%%%%%%%%%%%%%%%%%%%%%%%%%%%%%%%%%%%%%%%%%%
%\setcounter{section}{-1} %% Remove this when starting to work on the template.
\nolinenumbers

\section{Introduction}
To date, the LIGO and Virgo interferometric detectors~\cite{aLIGO2015,aVirgo2014,NardecchiaVirgo} have registered 90 gravitational wave (GW) events from the coalescence and merger of two compact bodies, either a pair of black holes (BHs) or neutron stars (NSs) or a mixed NS--BH system~\cite{GWTC-1,GWTC-2,GWTC-2.1,GWTC-3}. Each confirmed detection, starting from the first event in September 2015, has added a big piece of information about our comprehension of nature. However, there is still a huge pile of missing pieces to be added to the puzzle before having a complete picture of the GW sky. Indeed, the GW community is only at the beginning of the newborn GW astronomy. Gathering all the information collected so far, GW scientists have been able to prove, among other things, the existence of GWs, the possibility of having BHs in binaries, and the formation of heavy elements in NSs mergers, etc., just by considering the class of transient GW events. In principle, the number of potentially emitting sources is really large, since any system is able to generate GWs whenever there is a non-vanishing mass quadrupole moment.
In particular, other astrophysical systems are able to emit GWs potentially detectable by Earth-based interferometers such as LIGO, Virgo and KAGRA~\cite{aLIGO2015,aVirgo2014,KagraAkutsu2019}. In this review, I will focus on the discussion of the prospects of detection for a particular subset of GW signals called continuous waves (CWs). Although the emitted signal in this case might be difficult to catch in the data, a detection would certainly be of wide interest to the full scientific community. In Section~\ref{sec:phys}, I will provide some information about the physical systems that can emit CWs signals. In Section~\ref{sec:searches}, I will provide an up-to-date discussion of the data analysis techniques used for this type of problem. Finally, in Section~\ref{sec:results}, I will report the latest observational results, while conclusions are given in Section~\ref{sec:future}.

%%%%%%%%%%%%%%%%%%%%%%%%%%%%%%%%%%%%%%%%%%
\section{Sources of Continuous GWs}
\label{sec:phys}
The detection of continuous GW signals represents one of the next milestones to reach in gravitational-wave astronomy. 
Long-lasting signals, persistent over the full observing run (i.e., the data taking period), or in its shorter duration version, the so-called long CW transients, are typically emitted by fast spinning asymmetric NSs, both in accreting systems or as isolated sources. To this canonical set of sources, recent studies have reported the possibility of also observing  the continuous gravitational emission by potential dark matter (DM) candidates. Other potential sources of CWs are boson stars~\cite{DiGiovanniBosonstars2020,Sanchis-Gual2019} and Thorne-\.{Z}ytkow Objects~\cite{DeMarchiZytkow2021}, although no associated search in advanced detector data has been carried \mbox{out yet. }

In this section, I will review the main  emission models proposed in the literature. Previous reviews about CW signals and searches can be found in~\cite{TenorioCW2021,jonesCW2021,SieniawskaCW2019,GlampedakisNS2018,RilesCW2017,laskyNS2015} or in the more general GW reviews~\cite{BersanettiGW2021,CaudillGW2021}.

\subsection{Neutron Stars}
\label{ssec:NStheo}
The physical condition of NS matter is unique, making these fascinating compact objects precious laboratories for subatomic and fundamental physics~\cite{Lattimer2004,Haensel2007}.
High density and pressure are reached as the inner core is approached, with an average density comparable with terrestrial atomic nuclei $(\rho_0\sim 2.5\times 10^{14}~\rm g/cm^3)$, up to values one order of magnitude bigger in the inner core. Surface magnetic fields can reach values up to $10^{12}$~Gauss 
or $10^{15}$ Gauss for strongly magnetized NSs, also known as magnetars~\cite{Duncan1992,Esposito2020}.
Given the extreme condition of NS matter, the equation of state (EOS) is not completely known since superfluidity, superconductivity~\cite{Haskell2018,Chamel2017}, exotic states or quantum effects might be present~\cite{Baym2018,Oertel2017,Stone2021,Burgio2021}. 
The actual NS composition is still unknown. 
In particular, when higher densities are reached, 
the hadronic nuclear matter undergoes a de-confinement transition to a new phase of quarks and gluons~\cite{Annala2020}, although how this phase transition occurs is still unclear. In a different scenario, the NS inner core can be composed of hyperons~\cite{Vidana2018}, Bose--Einstein, pion or kaon condensates~\cite{pethick2015boseeinstein,BAYM1974,Ramos2001} and more. Furthermore, some of the above-mentioned states of matter can be simultaneously present inside the compact object.
The NS matter EOS directly reflects on the star's observable global parameters such as the mass and the radius. Theoretical and observational efforts, including those related to GWs,  to measure the NS masses and radii~are ongoing \cite{Ozel2016,Watts2016,Steiner2013,Lattimer2019,Llanes-Estrada2019}. To set constraints on the EOS of isolated NSs, both mass and radius should be measured, and some assumptions on the pulse profiles should be made (see, e.g.,~\cite{Bogdanov2019}). NS masses are typically measured from pulsar observations, while radius measurements are more challenging and strongly depend on the X-ray emission model assumed. To date, the NS measured masses lie in the range $[1.1;2.2]~M_{\odot}$, with the most massive source being the millisecond pulsar J0740+6620 with a mass of $2.14~M_{\odot}$~\cite{Cromartie2019}. Direct measurements of radii do not exist; however, estimates can be made, for instance, by combining the measured flux and temperature with the source distance. Current estimates report NS radii in the range  $[9;13]~\rm km$~\cite{Guillot2014,Nattila2017,Wei2019}. 
Further improvements on the NS radius and mass estimates are provided by the NICER mission~\cite{Gendreau2017NICER,Miller2019NICER}.
The detection of the first NS merger event GW170817 by LIGO and Virgo~\cite{Abbott2017GW170817}, followed by the electromagnetic (EM)  gamma-ray burst GRB170817~\cite{Abbott2017GRB170817A}, and the optical transient signal of a kilonova (AT 2017gfo)~\cite{Abbott2017multimessenger}, provided a new tool for NS mass and radius estimates~\cite{Abbott2018GW170817EOS}, extending the possible NS mass range. Other GW events involving potential NS have been reported to date: GW190425~\cite{Abbott2020GW190425} with a progenitor mass of $[1.12;2.52]~M_{\odot}$ and GW190814~\cite{Abbott2020GW190814} involving a compact object with a mass of $2.50-2.67~M_{\odot}$. No EM counterpart has been observed from the two events, suggesting that these are the smallest BHs to date, although the possibility of being the most massive NSs measured so far cannot be completely excluded. To this list of GW signals involving at least a (potential) NS, there are the events: GW190917\_114630~\cite{GWTC-2.1}, GW191219\_163120~\cite{GWTC-3}, GW200210\_092254~\cite{GWTC-3} and the two GW200105\_162426 and GW200115\_042309 discussed in~\cite{Abbott2NS2021,GWTC-3}.

The formation of an NS could happen through two main channels:  (i) core-collapse supernova explosions (CCSN), forming a very hot proto-neutron star, followed by a cooler stable NS~\cite{Cerda-Duran2018}, (ii) after a binary NS merger~\cite{Baiotti2017}. In these two scenarios, fast-rotating and highly magnetized young NS (in particular magnetars) could be formed.

Spinning NSs are expected to emit GWs continuously if they are asymmetric with respect to their rotation axis~\cite{haskell2021gravitational,Shapiro1983}. Potential targets for Earth-based detectors are galactic sources.
The signal produced as the star spins is almost monochromatic and with a frequency proportional to the star’s spin frequency.
Four different scenarios are considered possible for the observation of CWs, namely (i) very hot newborn NSs, right after the formation, i.e., after merger and core-collapse supernova, reaching rotating frequencies close to their Keplerian limit; (ii) young NSs such as stable supernova remnants during the initial spin-down phase; (iii) accreting low mass X-ray binaries (LMXBs) emitting thermal X-ray radiation and characterized by a small spin-up; (iv) fast-spinning pulsars, i.e., old millisecond pulsars with small spin-down.
Different CW emission models are considered (see~\cite{haskell2021gravitational}), and these include the emission due to the presence of non-axisymetries in rigid rotating triaxial bodies, rotating about one of its principal axes (equivalent to a biaxial rotor not spinning about one of the principal axes); oscillation of the star; or more complicated scenarios such as free precessing systems or heterogeneous stars (when multiple phases of component matter are simultaneously present, including superfluidity). 

If from one side, some of the GW emission mechanisms are well understood, the actual factors that cause the asymmetry in the star are still under debate.  Indeed, the asymmetry could be triggered by the presence of residual crustal deformations (e.g., after a fast cooling of the NS crust), the presence of a strong inner magnetic field not aligned with the star's rotation axis, or the presence of magnetic or thermal ``mountains''  (see~\cite{laskyNS2015} for a review). The maximum ellipticity (i.e., deformability) the star can sustain depends on both the NS EOS and the breaking strain of the crust~\cite{GlampedakisNS2018}.
I briefly review the two main CW emission mechanisms happening in NSs and how these reflect on the expected signal amplitude.

{\bf Ellipticity driven emission: %MDPI: OJP ISNECESSARY
} the GW amplitude for an NS with asymmetries can be expressed in terms of the product between the moment of inertia $I_{3}$ along the spin axis and the star's degree of asymmetry, also known as ellipticity $\varepsilon$:
\begin{equation}
\label{eq:amplitude}
h_{0}=\frac{16 \pi^{2} G}{c^{4}} \frac{I_{3} \varepsilon f_{\mathrm{spin}}^{2}}{r}
\end{equation}
where $f_{\mathrm{spin}}$ is the star spin frequency of a source at a distance $r$.  Most of the observed NSs have rotation frequencies below 1 kHz, i.e., in the sensitivity range of ground-based interferometers such as advanced LIGO/Virgo. 
The ellipticity is given by $\varepsilon=\frac{I_{2}-I_{1}}{I_{3}}$, assuming $I_{1}\neq I_{2}$.
Most of the parameters entering in Equation (\ref{eq:amplitude}) can be somehow constrained by astronomical observations, except for the ellipticity $\varepsilon$ for which only theoretical maxima estimates exist~\cite{Ushomirsky2002,Owen2005,Johnson-McDaniel2013,Woan2018,Gittins2020,Gittins2021}. Current estimates only report upper limits in the $10^{-7}$--$10^{-5}$ range, while an actual estimate of a fiducial ellipticity of ${\sim}10^{-9}$, possibly sustained by millisecond pulsars, is given in~\cite{Woan2018}. 

For the case of elastic stresses, the maximum sustainable ellipticity is related to the crustal breaking strain $u_{\rm break}$ as~\cite{Haskell2006}
\begin{equation}
\varepsilon<2 \times 10^{-5}\left(\frac{u_{\text {break}}}{0.1}\right)
\end{equation}
with estimated values of $u_{\text {break}}=0.1$, much larger than standard terrestrial materials~\cite{Horowitz2009}. However, these constraints strongly depend on the actual EOS (see, e.g.,~\cite{GlampedakisNS2018}). 

For quadrupolar deformations on spherically symmetric stars due to magnetic\linebreak \mbox{fields~\cite{Mastrano2011,Suvorov2016,Haskell2008}}, the ellipticity is given by the ratio of the magnetic energy and the GW energy. The actual deformation strongly depends on the magnetic field morphology. Indeed, poloidal fields will oblate the star, while toroidal fields tend to prolate the star. When the magnetic field is purely poloidal, the ellipticity will depend on the average $\bar{B}$ field as
\begin{equation}
\varepsilon \approx 10^{-12}\left(\frac{\bar{B}}{10^{12} \mathrm{G}}\right)^{2}.
\end{equation}

In the case %\textcolor{red}{(Please check whole tex and confirm if the noindent below equation is necessary?)}
of a fully toroidal magnetic field, the star's ellipticity is given by
\begin{equation}
\varepsilon \approx 10^{-11}\left(\frac{\bar{B}}{10^{12} \mathrm{G}}\right)^{2}.
\end{equation}

In a realistic scenario, also known as ``twisted-torus'' configuration, both the poloidal and toroidal components will contribute to the $B$ field (for a detailed discussion, see~\cite{haskell2021gravitational,laskyNS2015,GlampedakisNS2018}). 
Given the uncertainty connected to the estimation of $\varepsilon$ and the moment of inertia $I$, typically the deviation from axisymmetry is constrained in terms of the quadrupole  $Q_{22}$ (assuming that the $l=m=2$ mode dominates)~\cite{Owen2005}
\begin{equation}
   Q_{22}=\sqrt{\frac{8\pi}{15}} \varepsilon I.
\end{equation}

{\bf R-mode emission: %MDPI: Is the bold necessary? YES
} a different emission channel for CW radiation is given by the Rossby (r-)modes oscillations in rotating stars. R-mode emission is provided by the Chandrasekhar--Friedman--Schutz instability, leading to the rapid growth of the r-mode amplitude until a saturation amplitude is reached and the growth of the mode stops~\cite{Owen1998}. These amplitudes are damped by the viscosity of the star. Given these two opposite effects, where the amplitude increases with the spin of the star and is suppressed by the viscosity, dependent on the temperature, an instability window is defined in the angular velocity-temperature plane~\cite{Lindblom1998}.  Typically, NSs are unstable to r-modes except at extreme temperatures. At very low temperatures, the shear viscosity dominates the damping, while at very high temperatures, the bulk viscosity takes the lead in the r-mode suppression. Given this strict relation with the inner physics of the NS during the r-mode emission, constraints on the 
GW observable properties are interesting tools for the study of the EOS~\cite{haskell2021gravitational,Kokkotas2016}. A typical parameter constrained in CW searches for r-modes is the r-mode amplitude $\alpha$. Considering a source at a distance $r$ the GW strain $h_{0}$ is given by~\cite{Owen2010}:
\begin{equation}
h_{0}=\sqrt{\frac{2^9\pi^7}{5}} \frac{G}{c^{5}} \tilde{J} M R^{3}\frac{f_{\rm gw}^{3} \alpha}{r}
\end{equation}
where the dimensionless constant $\tilde{J}$ is connected to the source density and EOS~\cite{Owen1998,Owen2010}. $M$ and $R$ are the NS mass and radius. The GW frequency $f_{\rm gw}$ is proportional to the oscillation angular frequency $\omega=2\pi f_{\rm gw}$, which is related to the rotation angular velocity as  \mbox{$\omega{\sim}-4\Omega/ 3 $}. If relativistic corrections and other effects due to the rotation are considered~\cite{Lockitch2003,Idrisy2015}, e.g., those related to the EOS, the relation between the GW frequency and the star spin frequency can be parametrized as~\cite{Caride2019,Idrisy2015}:
\begin{equation}
f_{\rm gw}= A f_{\rm spin} + B \left(\frac{f_{\rm spin}}{f_{\rm Kepler}}\right)^2 f_{\rm spin}
\end{equation}
where $f_{\rm Kepler}$ is the Kepler frequency of the star, i.e., its  maximum sustainable spin frequency. The values for A and B lie in a range $1.39<A<1.57$ and $0<B<0.195$. 

Ellipticities are also expected to be present in LMXBs, due to an asymmetric accretion process, producing, for instance, thermal gradients in the crust~\cite{Ushomirsky2002}.
In these systems, a spinning NS is accreting matter from a companion, forming a circumstellar accretion disk around it and emitting strong X-ray radiation.
These systems represent interesting targets for CW searches~\cite{WattsLMXB2008} given that when the accretion torque balances the GW torque, LMXBs  become very stable GW emitters. In particular, the GW amplitude of an LMXBs is proportional to the square root of the X-ray flux. For this reason, very bright sources such as Scorpius X-1 (Sco X-1) or XTE J1751-305 represent ideal targets.  It has been observed that all the NSs in LMXB systems have maximum spinning frequencies below their Keplerian breakup limit. Indeed, the fastest LMXB observed has a spin frequency of ${\sim}700~\rm Hz$, while according to~\cite{Haskell2018sub}, the lowest breakup frequency allowed is ${\sim }1200~\rm Hz$, although this limit strongly depends on the assumptions made for the EOS of the star.
This suggests that this limit is strongly related to the GW torques~\cite{Haskell2018sub,Gittins2019} due to oscillations or asymmetries, although it could be due to the interaction of the accreting disk with the NS magnetic field~\cite{Andersson2005}. The search for CWs from accreting sources is complicated mainly by two factors~\cite{WattsLMXB2008}: (i) the source is spun up by the companion during the accretion, and a spin-wandering effect~\cite{Mukherjee2018} could be present; (ii) an additional Doppler effect  by the binary orbital motion is present~\cite{Leaci2017}.  
Eventually, variations in the  matter accretion rate directly reflect into irregular spinning frequency. To make these searches even more complicated, often the spin frequency is not known, as in the case of Sco X-1.

Different approaches are used, tailored to the emission model considered for the search. For instance, some methods are adapted and/or new algorithms have been developed for the case of a long duration transient emitted by postmerger remnants, or for the case of CWs from glitching pulsars. In the postmerger remnant case, the source is expected to have a high spin-down rate; hence, the frequency will strongly vary with time. In the glitching pulsar case, the frequency suddenly changes,  a sudden spin-up occurs, which may enhance the GW emission for a short time~\cite{Yim2020}, and then the star relaxes back to the standard spin-down scenario.

In thee case of detection, it is possible to extract some information about the main CW emission mechanism happening in an NS, by looking at the GW frequency measured (\mbox{see ~\cite{jonesCW2021}}). In particular, it is known that the GW frequency is related to the star spin frequency according to $f_{\rm gw}=k\left(f_{\mathrm{spin}}+f_{\mathrm{prec}}\right)$. Here, $k$ is a proportionality factor dependent on the model considered. In the simplest case of no precessing ($f_{\mathrm{prec}}=0$) rigid rotating bodies, with asymmetries supported by elastic and/or magnetic stresses, $k=2$, and hence, $f_{\rm gw}=2f_{\mathrm{spin}}$.
In r-mode oscillations scenarios, the estimated  emitted frequency  is $f_{\rm gw}{\sim}\frac{4}{3}f_{\mathrm{spin}}$, although deviations from this number are expected when relativistic corrections are included and/or different EOS are considered. In this case, the GW amplitude of the r-mode signal is parametrized in terms of the r-mode amplitude $\alpha$, and it explicitly depends on the mass and radius of the NS and on the star EOS.  For the case of CWs emitted by LMXB systems, accreting matter from a companion star, the relation between the spin frequency and the GW frequency is expected to be $f_{\rm gw}=2f_{\mathrm{spin}}$.

As the star spins, its angular momentum will be radiated away, e.g., via GW emission, decreasing its frequency as \endnote{Given the %MDPI: footer is not allowed, we change into notes, please confirm YES
proportionality between $f_{\rm gw}$ and $f_{\rm spin}$, here $f$ can indicate any of the two frequencies.}
\begin{equation}
\label{eq:powlaw}
    \dot{f}=K f^{n}
\end{equation}
where the negative constant $K$ and the braking index $n$ depend on the energy loss mechanism as $n=f \ddot{f} / \dot{f}^{2}$.
The spin-down process may be more complicated than the simple power law in Equation  (\ref{eq:powlaw}), and in general, it is possible that the energy loss is due to a simultaneous contribution from both the gravitational and magnetic dipole emission. For a spin-down fully dominated by a gravitational emission~\cite{Palomba2005}, i.e., a gravitar, $n=5$ (or $n=7$ if the GW emission happens via r-mode), while $n=3$ for magnetic dipole emission.  
An estimate of the age of the source can be expressed as
\begin{equation}
\label{Eq:age}
\tau=-\left[\frac{f}{(n-1) \dot{f}}\right]\left[1-\left(\frac{f}{f_{0}}\right)^{(n-1)}\right]
\end{equation}
here, $f_{0}$ is the star birth rotation frequency. For old sources (with $f \ll f_{0}$), the age is simplified as

\begin{equation}
\label{Eq:ageold}
\tau \approx-\left[\frac{f}{(n-1) \dot{f}}\right]
\end{equation}

More complicated emission scenarios are described in the literature. For instance, when the star is in free precession, dual harmonic emissions can be present at once and twice the  star's spinning frequency. %Please confirm intended meaning is retained-> changed from  can be present at the star's spinning frequency or twice that frequency.
A particular case of dual harmonic emission is encountered when the star has a superfluid core pinned to the crust and not aligned with the principal axes of the moment of inertia~\cite{Jones2010}.
More complex emission models exist, e.g., when the configuration is that of a triaxial freely precessing body or for the case of a deferentially rotating two-component star.  In this case, the emission may happen on more than two harmonics, complicating the signal waveform to be searched for. Typically these models are not considered in real searches.

The class of CWs signals is then a good benchmark to study the state of matter at higher densities. In addition to this aspect, CW detections can be used to test general relativity (GR). According to GR, only two tensor polarization states exist for GWs. When more generic metric theories of gravity are assumed, the number of polarizations increases up to six, allowing for two scalar and two additional vector  modes. In particular, for the Brans–Dicke theory of gravity~\cite{Brans1961}, only one additional scalar polarization state is predicted to be present along with the two tensor polarizations predicted by GR.
As discussed \mbox{in~\cite{Isi2015,Isi2017,Verma2021}}, it is possible to put constraints on these extra polarizations using CWs detections from known pulsars. Indeed, the detector response function to a given polarization depends on the direction of propagation of the GW along the two detector arms (see, e.g.,~\cite{Schutz_2011}). Given the long duration nature of CWs, different antenna patterns will produce a different sidereal amplitude modulation. This means that all the relevant information to distinguish between the polarizations is encoded in the sidereal-day-period amplitude modulation of the signal. This is true only if the signal phase evolution for non-GR polarizations is equal to that assumed for GR. 

\subsubsection{Strain Amplitude Limits}
\label{sssec:limitstheo}
It is possible to identify promising targets for CWs by computing some limits based on the information about the source. These quantities constraint the maximum GW emission of a given source. Three main limits are used in CW searches: the spin-down limit, the age-based limit and the torque balance limit. 

\emph{Spin-down limit:} %MDPI:\textcolor{red}{( Is the italics necessary?)} YES
 generally, for known pulsars with precisely measured $f_{\rm spin}$ and $\dot{f}_{\rm spin}$, it is possible to define the so-called spin-down limit on the maximum detectable strain~\cite{Shapiro1983}. If all the energy loss during the spin-down is released in the form of GWs, by equating the GW power emission to the time derivative of the rotational kinetic energy, one obtains a source at a distance $r$:
\begin{equation}
\label{eq:spindownlimit}
\begin{aligned}
h^{\rm sd}_{0} &=\frac{1}{r} \sqrt{\frac{5}{2} \frac{G}{c^{3}} I_{\mathrm{3}} \frac{\left|\dot{f}_{\mathrm{gw}}\right|}{f_{\mathrm{gw}}}} =2.5 \times 10^{-25}\left(\frac{1 \mathrm{kpc}}{r}\right) \sqrt{\left(\frac{1 \mathrm{kHz}}{f_{\mathrm{gw}}}\right)\left(\frac{\left|\dot{f}_{\mathrm{gw}}\right|}{10^{-10} \mathrm{~Hz} / \mathrm{s}}\right)\left(\frac{I_{\mathrm{3}}}{I_{0}}\right)} .
\end{aligned}
\end{equation}
where $I_{0}=10^{38} \mathrm{~kg} \mathrm{~m}^{2}$. 
This gives an absolute upper limit to the amplitude of the CW signal emitted by a pulsar.

\emph{Age-based limit:} an alternative way to check the goodness of a potential CW target, when the rotational parameters are unknown, is to look at the so-called indirect age-based limit. Similar to the spin-down limit, assuming that all the energy lost is radiated away with GWs, it is possible to express the spin-down limits in terms of the age of the star $\tau$ using Equation  (\ref{Eq:age}) and a braking index $n=5$~\cite{Wette2008}
\begin{equation}
\label{eq:agebasedlimit}
h_{0}^{\text {age }}=2.27 \times 10^{-24}\left(\frac{1 \mathrm{kpc}}{r}\right)\left(\frac{1 \mathrm{kyr}}{\tau}\right)^{1 / 2}\left(\frac{I_{3}}{I_{0}}\right)^{1 / 2}.
\end{equation}

\emph{Torque balance limit:}
empirical upper limits can also be defined  for accreting binaries. Assuming that the GW emission is completely balanced by the angular momentum added via accretion, the so-called torque-balance limit can be derived~\cite{Bildsten1997}. Assuming that the accretion luminosity is fully radiated as X-rays,  this provides an estimate on the mass accumulation rate; hence, the limit can be written in terms of the observable X-ray flux $F_{X}$ and the star spin frequency (GW frequency):
\vspace{12pt}
\begin{equation}
\label{eq:torquebalancelimit}
h^{\rm tb}_{0}=5 \times 10^{-27} \left(\frac{F_{X}}{F_{\star}} \right)^{1 / 2}\left(\frac{R}{10 \mathrm{~km}}\right)^{1 / 2}\left(\frac{r_{m}}{10 \mathrm{~km}}\right)^{1 / 4}\left(\frac{1.4 \mathrm{M}_{\odot}}{M}\right)^{1 / 4}\left(\frac{700 \mathrm{~Hz}}{f_{\rm gw}}\right)^{1 / 2}
\end{equation}
with $F_{\star}=10^{-8} \operatorname{erg} \mathrm{cm}^{-2} \mathrm{~s}^{-1}$. $M$ and $R$ are the NS mass and radius and $r_{m}$ is the lever arm, usually equal to $R$ or to the Alfven radius, i.e., the  radius corresponding to the inner edge of the accretion disk.

%%%%%%%%%%%%%%%%%%%%%%%%%%%%%%%%%%%%%%%%%%

\subsection{Dark Matter Candidates}
\label{ssec:DMtheory}
A CW-like signature is expected to be produced in the data also by several potential DM candidates~\cite{Bertone:2016nfn}. For a few years, interferometric detector data have been used to look for evidence of DM, both in terms of GW, emitted by systems made up of particles from the dark sector, and as direct DM detectors, assuming a DM coupling to the detector components.  For a review of all the DM candidates that can be investigated using GWs detectors, see~\cite{Bertone_2020}. Potential DM candidates may be ultra-light bosons, including dark photons or quantum-chromo-dynamic axions~\cite{Arvanitaki2010,Arvanitaki2011,Brito_2020,Peccei:1977hh}.
In this section, I review three main scenarios involving DM candidates, where a persistent CW-like signal is expected to be produced and actively searched by the CW community: boson clouds around spinning BHs, vector bosons in form of dark photons, compact dark objects or primordial black holes.  The latest observational results will be discussed in Section \ref{ssec:resultsunknown}.

\subsubsection{Boson Clouds}
Ultralight bosons condensates can clump around spinning BHs through superradiance~\cite{Arvanitaki2015,Arvanitaki2011,Brito_2020}. 
The BH-boson cloud system can be described using an analogy with the hydrogen atom model when the axion's Compton wavelength is comparable to the size of a BH, turning this system into a "gravitational atom".  %MDPI: removed italics
When boson fields are present nearby a spinning BH, these can interact with the BH and condensate around it, all occupying the same (quantum) state and reaching huge occupation numbers after exponential growth. This process takes place at the expense of the BH angular momentum, and indeed, the axion or axion-like particles scatter on the BH decreasing its spin. When the BH spin is low enough, the superradiance process
stops, and the cloud dissipates via axion annihilation into gravitons. During this depletion phase or when transitioning between levels of the gravitational atom, the system generates a quasi-monochromatic and long-duration signal that can be searched with CW methods. It should be noted, however, that the GW contribution from transition levels is sub-dominant. 
The most interesting scenario for Earth-based detectors such as LIGO, Virgo and KAGRA is the annihilation process. Indeed, the signal produced in this case is in the detector sensitivity band for boson masses in the $10^{-13}-\times10^{-12}~\rm eV$ range.
If the boson self-interaction is negligible, the gravitational-wave signal frequency depends mainly on the mass of the boson and weakly on the mass
and spin of the BH. When self-interaction is considered, the approximation of the signal with a CW-like  shape is no longer valid since the signal is strongly suppressed~\cite{Baryakhtar2021}. In particular, if the self-interaction is stronger than the gravitational binding energy, the cloud becomes unstable, and it collapses in a process known as ``bosenova''. During this process, the system may produce a burst  of GWs~\cite{Arvanitaki2011}.
In general, the annihilation signal cannot be fully characterized  unless some  assumptions on the BH are made. Indeed, the emitted signal depends on the mass, spin, distance, and age of the BH.

Let us briefly review the main characteristic of the signal emitted by BH-boson clouds as in~\cite{Palomba2019PRL,Zhu2020boson,Arvanitaki2010,Arvanitaki2011,Arvanitaki2015}. 
When the superradiant condition is satisfied, i.e., when the boson angular frequency is less than the BH's outer horizon angular frequency, the scalar field starts to clump around the BH. This effect is maximized when  the particle's reduced Compton wavelength $\lambda_{b}=\frac{\hbar c}{m_{b}}$, which depends on the boson mass $m_{b}$, is comparable to the BH Schwarzschild radius $R=\frac{2G M_{\mathrm{BH}}}{c^{2}}$. This superradiant instability phase has a typical duration of
\begin{equation}
\tau_{\text {inst }} \approx 20\left(\frac{M_{\mathrm{BH}}}{10~M_{\odot}}\right)\left(\frac{\alpha}{0.1}\right)^{-9}\left(\frac{1}{\chi_{i}}\right) \text { days, }
\end{equation}
dependent on the BH mass $M_{\rm BH}$ and dimensionless spin $\chi_{i}$. Returning to the analogy with the hydrogen atom, $\alpha$ is the fine-structure constant in the gravitational atom 
\begin{equation}
\alpha=\frac{G M_{\mathrm{BH}}}{c^{3}} \frac{m_{b}}{\hbar}.
\end{equation}
As long as the BH spin is above the critical spin $\chi_{c} \approx \frac{4 \alpha}{1+4 \alpha^{2}}$, the cloud will continue to grow and the BH will decrease its spin. 
Once equilibrium has been reached, the boson cloud mass is dissipated 
on a timescale\endnote{for $m = 1$ and for $\alpha \ll 0.1$.}
\begin{equation}
\tau_{\mathrm{gw}} \approx 6.5 \times 10^{4}\left(\frac{M_{\mathrm{BH}}}{10 ~M_{\odot}}\right)\left(\frac{\alpha}{0.1}\right)^{-15}\left(\frac{1}{\chi_{i}}\right) \text { years. }
\end{equation}

The GW amplitude decays in time as $\left(1+t / \tau_{\mathrm{gw}}\right)^{-1}$ from a starting strain amplitude determined mainly by the BH  and boson masses (entering in the definition of $\alpha$) and the BH initial spin 
\begin{equation}
h_{0} \approx 3 \times 10^{-24}\left(\frac{\alpha}{0.1}\right)^{7}\left(\frac{\chi_{i}-\chi_{c}}{0.5}\right)\left(\frac{M_{\mathrm{BH}}}{10~\mathrm{M}_{\odot}}\right)\left(\frac{1~\mathrm{kpc}}{r}\right).
\end{equation}
During this emission phase, the GW  emitted frequency, which is twice the frequency of the field, is also dependent on the two-component masses as
\begin{equation}
    f_{\mathrm{gw}}  \simeq 483 \mathrm{~Hz}\left(\frac{m_{\mathrm{b}}}{10^{-12} \mathrm{~eV}}\right) \left[1-7 \times 10^{-4}\left(\frac{M_{\mathrm{BH}}}{10~M_{\odot}} \frac{m_{\mathrm{b}}}{10^{-12}~\mathrm{eV}}\right)^{2}\right]
\end{equation}
During the depletion phase, a small spin-up is present due to the loss of mass. Indeed, during the cloud formation phase, a large spin-down rate is present, although the expected signal strain is still too low to be detected. This drift changes when the cloud starts to evaporate. The actual spin-up rate during the depletion phase is strongly dependent on the boson self-interaction constant. For simplicity, we consider the case of a negligible self-interaction when the spin-up due to annihilation is the dominant drift:
\begin{equation}
\dot{f}_{\mathrm{gw}} \approx 7 \times 10^{-15}\left(\frac{m_{\mathrm{b}}}{10^{-12} \mathrm{~eV}}\right)^{2}\left(\frac{\alpha}{0.1}\right)^{17} \mathrm{~Hz} / \mathrm{s}.
\end{equation}
When the self-interaction starts to be non-negligible, the spin-up rate is enhanced by two contributions, one from the bosons energy level transition and a second from the change in the self-interaction energy as the cloud depletes. At this point, the signal is expected to be shorter or with a smaller amplitude.
For a more general discussion on the role of the boson self-interaction parameter, see~\cite{Baryakhtar2021}. 
In current CW searches, the main contribution to the GW emission considered for these models are those from the main growing state since the second fastest growing state emission is weaker than the first level one and is not \mbox{fully quadrupolar.}

Analogous to the standard CW case for NSs, Doppler modulations should be considered when looking for the signal in the detector data (see Section %MDPI: Newly added information, please confirm.
  \ref{ssec:resultsunknown} for the latest observing results from this field).
Finally, we notice that the same superradiance effect described above is somehow expected also for rotating NS~\cite{Cardoso2017,Day2019}, although it is not clear if the GW emission is loud enough to be detected by currently operating detectors.

\subsubsection{Ultralight Vector Bosons: Dark Photons.}
DM particles are expected to interact with other DM particles in the same way as standard matter interacts with the EM force mediated by the photon. The equivalent force mediator for the dark sector is the dark photon. According to current theories, the dark photon is not expected to couple directly to standard model particles, although a small mixing-induced coupling to EM currents can be present~\cite{graham2021searches,Fabbrichesi2021}. 
Here, we focus on an ultralight DM candidate,  which behaves as a classical field interacting coherently with the atoms of the test masses. 
In particular, we assume that the dark photon is a vector boson coupled to the baryon or baryons minus leptons number $U(1)_B/U(1)_{B-L}$. In practice, if sufficiently light, the dark photon has a high phase-space density; hence, it behaves as a coherently oscillating classic field producing an oscillating force on dark charged objects~\cite{Pierce2018,Graham2016,Carney2021}. The same type of oscillation is expected for the tensor boson case~\cite{Armaleo2021}. The mass of the dark photon could be provided either by a dark Higgs boson or via the Stuckelberg mechanism~\cite{Raggi2015,graham2021searches,Fabbrichesi2021}. 
Different production mechanisms have been proposed for the dark photon, such as the misalignment mechanism associated with the inflationary epoch, light scalar decay or cosmic strings~\cite{Co2018,Co2019}.
Several attempts to set constraints on this type of DM are present in the literature~\cite{Graham2016,Guo2019Gaia,xue2021highprecision}.
The dark photon DM field oscillations act as a time-dependent equivalence principle-violating force acting on the test masses and producing a change in the relative length of the detector's arms. There exist two main signatures when test bodies interact with the dark photon  fields: a spatial gradient is present, producing a relative acceleration between the objects due to the different field amplitude; given the equivalence principle-violation of this force, test masses of different materials will experience different accelerations. 
For all Earth-based detectors, an additional effect due to the finite light travel time should be considered~\cite{Arvanitaki2015dilaton,Morisaki2021}.

I briefly review the main characteristics of the expected signal from dark photons while search results are discussed in Section~\ref{ssec:resultsunknown}. 
The Lagrangian of the massive vector field $A^{\mu}$, which couples to $B$ or $B-L$ current $J_{D}^{\mu}(D=B$ or $B-L$ ) as DM is given (in natural units) by 
\begin{equation}
\mathcal{L}=-\frac{1}{4} F^{\mu \nu} F_{\mu \nu}+\frac{1}{2} m_{A}^{2} A^{\mu} A_{\mu}-\epsilon_{D} e J_{D}^{\mu} A_{\mu},
\end{equation}
where $F_{\mu \nu}=\partial_{\mu} A_{\nu}-\partial_{\nu} A_{\mu}, m_{A}$ is the mass of the vector field, and $\epsilon_{D}$ is the coupling constant normalized to the EM one.
The dark photon  field $A_{\mu}$ can be approximated as a plane wave with a characteristic momentum  $\vec{k} \simeq m_{A} \vec{v_{0}}/\hbar$, within a coherence time $T_{\mathrm{coh}}\simeq 2 \pi \hbar (m_{A} v_{0}^{2})^{-1}$.
The local amplitude $A_{\mu, 0}$ of the dark electric gauge field is obtained assuming an energy density, $\frac{1}{2} m_{A}^{2} A_{\mu, 0} A_{0}^{\mu}$, equal to that of the local DM $\rho_{\mathrm{DM}}=0.4~\mathrm{GeV} / \mathrm{cm}^{3}$. 
The time-dependent force acting on the test masses produces a strain oscillating at the same frequency and phase as the DM field~\cite{Pierce2018,Guo2019O1,Morisaki2021,Michimura2020}. It should be noted that a similar signal may be produced in the case of scalar dark matter particles directly interacting with the mirrors. In particular, as described by~\cite{vermeulen2021}, the interaction of the scalar bosons with the beam-splitter will induce oscillations in the thickness of the mirror due to the oscillations of the fundamental constants~\cite{Stadnik2016,Stadnik2015,Grote2019}.
Considering only vector bosons in a non relativistic scenario, the contribution from the magnetic dark field is negligible if compared to the electric dark field\endnote{Hence, the time derivative of the time component of $A_{\mu}$ is negligible relative to $\vec{A}$.}. The dark photon DM background field will generate an acceleration on each $i$-th test mass as:
\begin{equation}
\label{Eq:accDP}
\vec{a}_{i}\left(t, \vec{x}_{i}\right)=\frac{\vec{F}_{i}\left(t, \overrightarrow{x_{i}}\right)}{M_{i}} \simeq \epsilon_{D} e \frac{q_{D, i}}{M_{i}} \partial_{t} \vec{A}\left(t, \overrightarrow{x_{i}}\right)
\end{equation}
$M_{i}$  and  $q_{D, i}$  are the total mass and dark charge of the test mass located at $x_{i}$.

This acceleration will cause a tiny differential relative
displacement between pairs of test masses that can be converted into a strain by integrating Equation  (\ref{Eq:accDP}) twice over time. Following~\cite{Morisaki2021,LVK2021O3DPDM}, the GW strain will have two contributes, one due to the relative variation of the arm length due to the dark force acting on the mirrors $h_{D}(t)$, and a second one due to the finite light-traveling time in the arm $h_{C}(t)$. These two contributions can be averaged over random polarization and propagation directions, and the resulting strain in SI units can be written as:
\begin{equation}
%\sqrt{\left\langle h_{D}^{2}\right\rangle} =C \frac{q}{M}\frac{e\epsilon_{D}}{2\pi c^{2}} \sqrt{\frac{2 \rho_{\mathrm{DM}}}{\epsilon_{0}}}\frac{v_{0}}{f_{0}}
\sqrt{\left\langle h_{D}^{2}\right\rangle} =C\frac{q}{M}\frac{e\epsilon_{D}}{2\pi c^{2}} \sqrt{\frac{2\rho_{\mathrm{DM}}}{\epsilon_{0}}}\frac{v_{0}}{f_{0}}=6.28\times10^{-27}\left(\frac{\epsilon_{D}}{10^{-23}}\right)\left(\frac{100 \mathrm{~Hz}}{f_{0}}\right)
%\frac{\hbar e}{c^{4} \sqrt{\epsilon_{0}}} \sqrt{2 \rho_{\mathrm{DM}}} v_{0} \frac{\epsilon}{f_{0}}  \simeq 6.56 \times 10^{-26}\left(\frac{\epsilon}{10^{-22}}\right)\left(\frac{100 \mathrm{~Hz}}{f_{0}}\right)
\end{equation}

\begin{equation}
\sqrt{\left\langle h_{C}^{2}\right\rangle} =\frac{\sqrt{3}}{2} \sqrt{\left\langle h_{D}^{2}\right\rangle} \left(\frac{2 \pi f_{0} L}{v_{0}}\right)  \simeq 6.21 \times 10^{-26}\left(\frac{\epsilon_{D}}{10^{-23}}\right).
\end{equation}
The total strain will be the root sum of the squares of the two contributions. The quantity $C$ is a geometric factor dependent on the position and orientation of the interferometer with respect to the DM wind.
The signal frequency is then determined by the dark photon mass $f_{0}=m_{A}c^2/(2\pi\hbar)$, corresponding for Earth-based detectors to dark photon masses in the range $10^{-14}$--$10^{-11} \mathrm{~eV}/c^2$. %MDPI: We change minus sign into endash please confirm -> OK
Given that the velocity of dark photon  particles follows a Maxwell--Boltzmann distribution, and assuming a virial velocity of $v_0=220~\rm km/s$, the signal frequency will lie in the $(f_0,f_0+\Delta f)$ range, where:
\begin{equation}
\frac{\Delta f}{f_0}=\frac{1}{2}\frac{v_{0}^{2}}{c^2}\approx 2.94 \times 10^{-7}.
\end{equation}

Furthermore, for the dark photon DM case, the Doppler shift due to Earth's rotation is present, although the frequency shift due to this effect is an order of magnitude smaller than the one from the Maxwell--Boltzmann spreading.

\subsubsection{Compact Dark Objects and Primordial Black Holes} 
Dark matter can be present in our Universe in the form of macroscopic objects, which we generically refer to as compact dark objects (CDOs). The actual formation channel and the origin of CDOs are still widely debated. According to some hypotheses, CDOs can be also PBHs. CDOs can form pairs and emit an almost monochromatic signal while being far from the coalescence phase.  
One option for the formation of the pair, proposed in~\cite{Horowitz2020CDO}, is to assume that these objects can be trapped inside normal matter via non-gravitational interactions, for instance, the Sun or even the Earth~\cite{Horowitz2020Earth}.  If  a second CDO is also trapped inside the same object, these can form a binary CDO emitting GWs with a long lifetime for low mass objects. According to~\cite{Horowitz2020CDO}, the probability of collisions between CDOs and the Sun is not negligible for CDOs masses ${\sim}10^{-10}M_{\odot}$; hence, there is a non-negligible possibility of having these pairs of CDOs formed in our solar system. 
Furthermore, PBHs can be constituents of a fraction of the DM in the Universe depending on their formation channels (see~\cite{carr2021primordial,carr2021primordialconst,Green2021primordial, carr2020primordial} for the latest reviews about PBHs as DM candidates). 
In particular, the origin of most of the sub-solar and planetary mass BHs is likely primordial, although speculations on alternative formation channels exist, e.g., created from NSs by the accumulation of DM and subsequent collapse into a BH~\cite{Kouvaris2018}.
For the context of CW signals, PBHs, as well as more generic CDOs, represent potential targets. Indeed, the GW signals emitted by these systems, when the two compact objects are far away from the coalescence and/or their masses are small enough, are modeled as quasi-monochromatic. This approximation is valid until the inspiral orbit does not reach the inner-most stable orbit. When the low-mass/far-from-coalescence conditions are dropped, the frequency evolution of these systems is better represented by a power law with a spin-up.  %Please confirm intended meaning is retained. -> Ican't see a diff

For the masses considered (${\leq}10^{-5}M_{\odot}$) in searches for CW signals from a pair of compact objects, the linear approximation is always valid  considering the frequencies of the detector sensitivity band. 
In addition, if the chirp mass is small enough, the frequency can be modeled exactly as the linear frequency Taylor expansion used for standard CW searches. This means that, for instance, a pair of inspiralling PBHs with chirp masses below $10^{-5}~M_{\odot}$ would emit a GW signal indistinguishable from those arising from non-axisymmetric rotating NSs spinning up.

Let us now consider the type of emission that could be released by these systems. The two systems of the two orbiting PBHs/CDOs can be approximated as a two-body problem of point mass objects in a circular orbit. The system loses energy as the inspiralling continues and the two masses approach each other. The signal model discussed in this section is widely described in the classical handbook of GW science~\cite{Misner1973GW,maggiore2008gravitational,Creighton2011GW} as well as in~\cite{Peters1964,Horowitz2020CDO,Tiwari2015,Miller2021PBH}.
The expected signal strain amplitude is that of a circular BH binary inspiral at a distance $r$ identified by a chirp mass $\mathcal{M}$ with a time to the coalescence $\tau=t_{c}-t$ :
\vspace{6pt}
\begin{equation}
h_{0}(t)=\frac{4c}{r}\left(\frac{G\mathcal{M}}{c^3}\right)^{5/3}\left( \pi f_{\mathrm{gw}}(t)\right)^{2/3}.
\label{eq:h0PBH}
\end{equation}
The amplitude is time-dependent and scales with the GW frequency, equal to twice the system orbital frequency. 
The GW frequency emitted is dependent on the system chirp mass $\mathcal{M}$  as
\begin{equation}
f_{\mathrm{gw}}(\tau)=\frac{(5 / \tau)^{\frac{3}{8}}}{8 \pi}\left(\frac{G \mathcal{M}}{c^3}\right)^{-5 / 8}.
\label{eq:fPBH}
\end{equation}
In this case, the expected signal frequency variation (the spin-up) due to the change in distance of the inspiralling system is given by:

\begin{equation}
\label{eq:spinupPBH}
\dot{f}_{\mathrm{gw}}=\frac{96}{5} \pi^{8 / 3}\left(\frac{G \mathcal{M}}{c^{3}}\right)^{5 / 3} f_{\mathrm{gw}}^{11 / 3}.
\end{equation}
Equation  (\ref{eq:fPBH}) is derived by integrating the expression for the spin-up. 
For low masses,  the frequency evolution can be simplified, assuming that the observing time $T_{\rm obs}$ is bigger than $\tau$ and assuming a small chirp mass. This means that the PBH/CDO spin-up rate $\dot{f}_{\mathrm{gw}}$ can be treated as a constant. In this case, the expected signal can be linearized as $f_{\mathrm{gw}}(t) \sim f_{0}+\left(t-t_{0}\right) \dot{f}_{\mathrm{gw}}$,
which coincides with the standard NS case (see Equation (\ref{eq:phase})).  On the other hand, if the spin-up is too high, this approximation is no longer valid, and the frequency evolution should be modeled as a power-law, as discussed in~\cite{Miller2021PBH}.
Furthermore, for this system, the signal undergoes a daily modulation at the detector, which should be considered in the analyses.

\section{Searches for CWs Signals with Earth-Based Detectors}
\label{sec:searches}
Different methods for the search for CW signals have been developed. The variety of methods reflects the different ways it is possible to look for these long-lasting signals, giving priority to the sensitivity of the pipeline (i.e., the search algorithm) or its robustness with respect to deviations of the signal from the assumed model.  

\subsection{The Signal at the Detector}
\label{ssec:signaldet}
CWs are by definition all those quasi-monochromatic GW signals characterized by a long duration, ranging from hours to years, and deeply buried in detector noise. When the CW signal reaches the detector, several modulation effects occur. The strain measured at the detector for a triaxial rotor is given by
\begin{equation}
\label{eq:detstrain}
%h(t)=F_{+}(t, \alpha, \delta, \psi)h_{+}(t)+F_{\times}(t, \alpha, \delta, \psi)h_{\times}(t)
h(t)=h_{0}\left[F_{+}(t, \alpha, \delta, \psi)\frac{\left(1+\cos ^{2} \iota\right)}{2} \cos \Phi(t)+F_{\times}(t, \alpha, \delta, \psi) \cos \iota \sin \Phi(t)\right]
\end{equation}
here, $(\alpha,\delta)$ are the right ascension and declination
of the source in the sky,  $\psi$  is the polarization angle, $F_{+/\times}$ are the detector response to the two $h_{+/\times}$ polarizations of the GW. The signal strain also depends on the inclination of the rotation axis to the line of sight $\iota$ and the GW phase $\Phi(t)$ (see, e.g.,~\cite{maggiore2008gravitational}). 
The phase of the GW signal is related to the frequency components $(f,\dot{f},\ddot{f},\ldots )$ at a given reference time $t_{0}$ as 
\begin{equation}
\label{eq:phase}
\Phi(t)=\phi_{o}+2 \pi\left[f\left(t-t_{0}\right)+\frac{\dot{f}}{2 !}\left(t-t_{0}\right)^{2}+\ldots \right]
\end{equation}
where $\phi_{0}$ is an initial phase. The signal phase is modulated mainly by the Doppler due to relative motion between the detectors and the source and other relativistic effects, namely the Einstein and Shapiro delays. The frequency at the detector will be spread with respect to the emitted frequency as: 
\begin{equation}
\label{eq:doppler}
f_{\rm gw}(t)=\frac{1}{2 \pi} \frac{d \Phi(t)}{d t}=f_{0}(t)\left(1+\frac{\vec{v}(t) \cdot \widehat{n}}{c}\right)
\end{equation}
where $\vec{v}$ is the velocity vector of the detector, while $\widehat{n}$ is the unit vector pointing to the source direction, both expressed in the Solar System Barycenter reference frame. Here, $f_{0}(t)$ is the GW frequency arriving at the detector at the time $t$ and emitted by the star at the reference time $t_0$.   As mentioned in the previous sections, $f_{0}(t)$ can be described by a Taylor expansion such as $f_{0}(t) \simeq f_0 + \dot{f}(t-t_0) + \dots$, where $\dot{f}$ is the first order spin-down parameter.

The phase modulation due to the Doppler Effect, apart from an irrelevant constant, can be obtained by integrating Equation  \eqref{eq:doppler} as:
\begin{equation}
\label{eq:doppphase}
\phi_{d}(t)=2 \pi \int_{t_{0}}^{t} f_{0}\left(t^{\prime}\right) \frac{\vec{v} \cdot \hat{n}}{c} d t^{\prime} \approx \frac{2 \pi}{c} p_{\hat{n}}(t) f_{0}(t)
\end{equation}
where $p_{\hat{n}}(t)$ is the position of the detector projected along $\hat{n}$.

\begin{figure}[H]
%\centering
\includegraphics[scale=0.32]{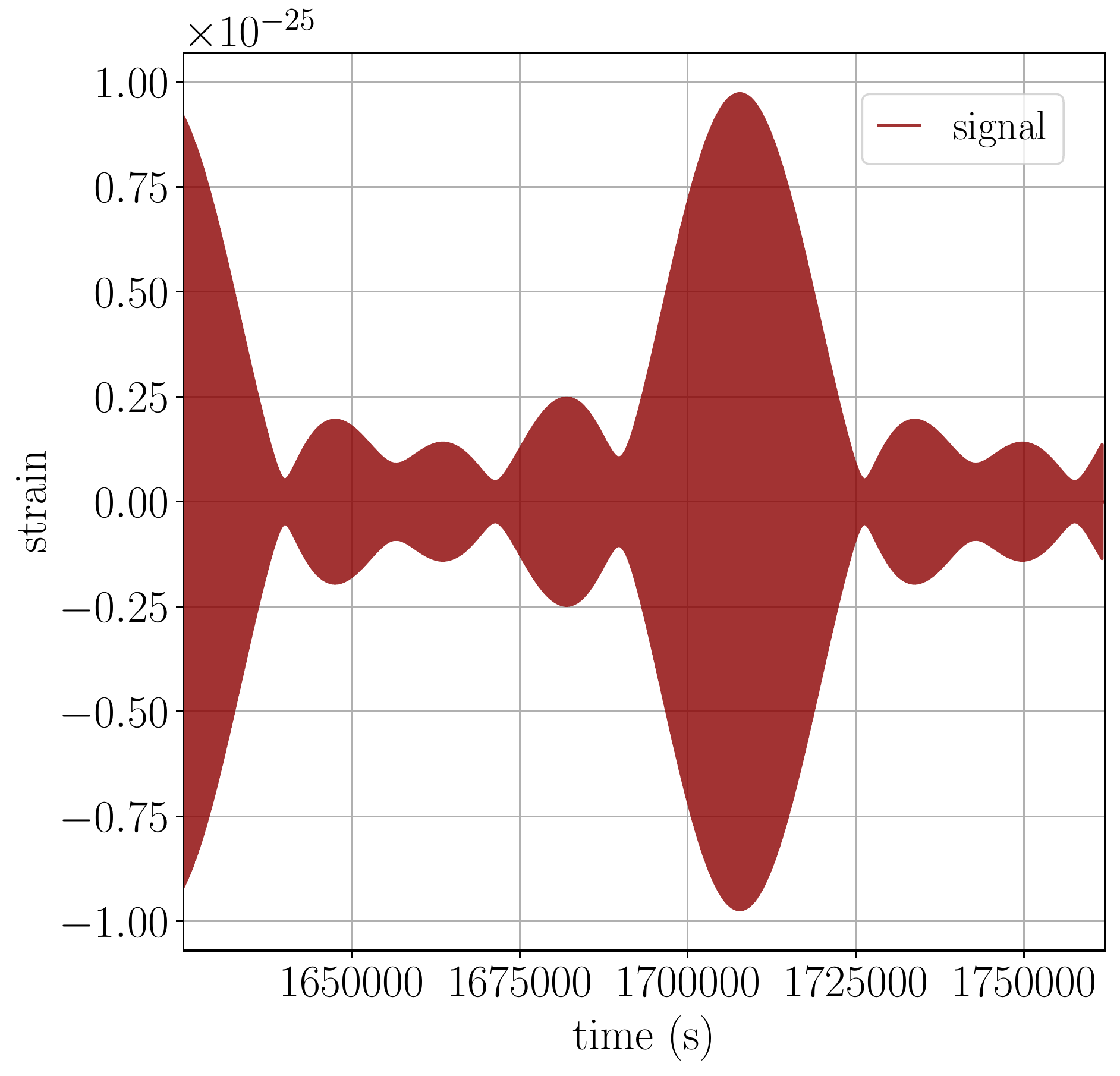}
\hspace{1cm}
\includegraphics[scale=0.32]{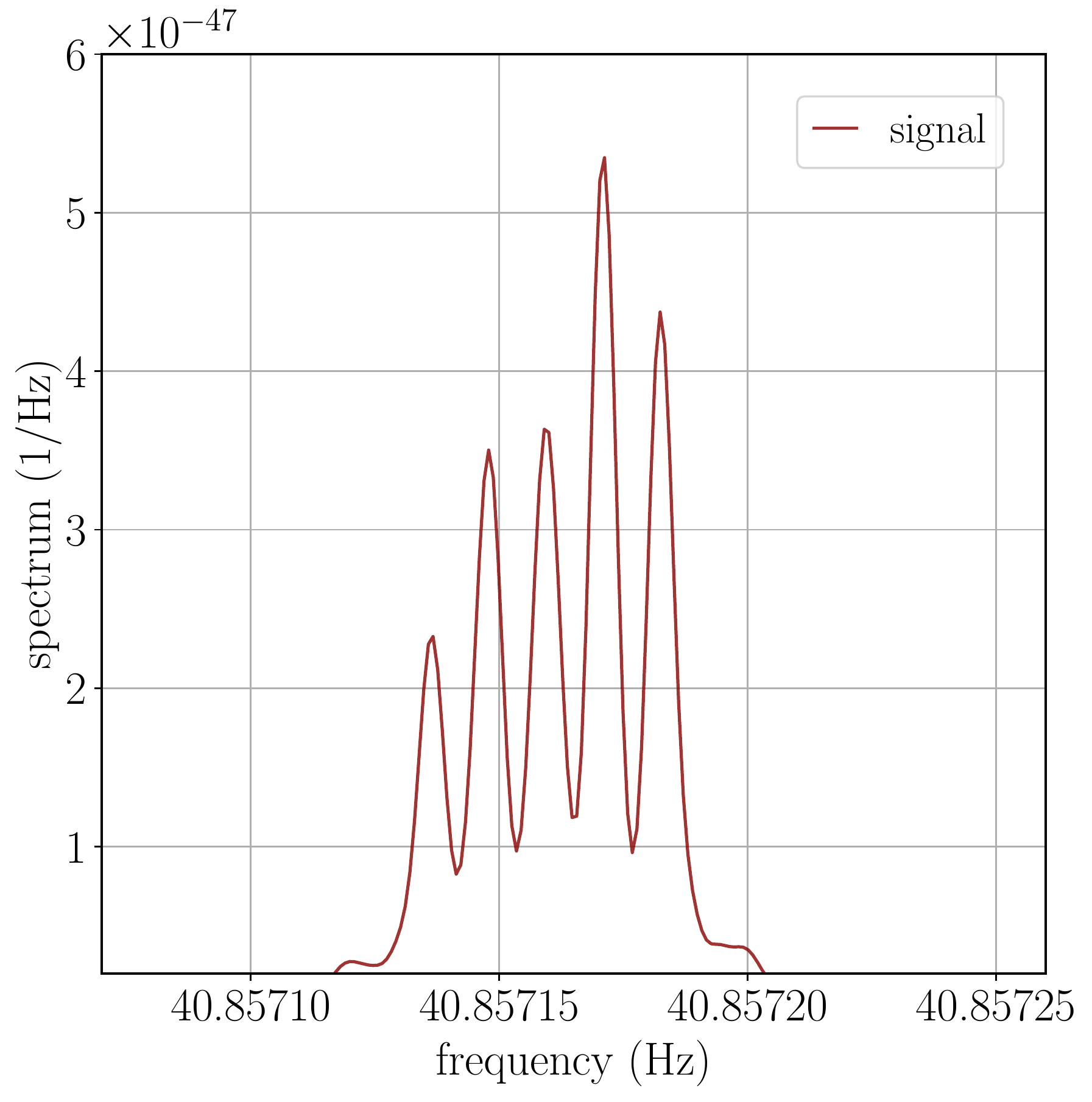}
\caption{Features 
of a CW signal without noise. (\textbf{Left}): time series of the signal before removing any modulation, the signal is modulated in amplitude and frequency. (\textbf{Right}): power spectrum after removing the frequency modulations, the five peaks due to the sidereal modulation are clearly visible.}
%\label{fig:strainpow}
\label{fig:strain}
\end{figure}

The pure signal strain at the detector, assuming the ideal situation where no noise is present, is depicted in Figure \ref{fig:strain}. 
The figure on the left reproduces a signal as it is received at the detector, and it is visibly modulated both in frequency and in amplitude. Even after the removal of the main modulation effects, namely the Doppler and the spin-down of the source,  the power spectrum reported on the right side is very different from the one produced by a purely sinusoidal signal. This amplitude modulation is produced by the antenna sidereal response; hence, by the time dependence of $F_{+/\times}$. In the right plot, five peaks are visible, and those are due to the Earth's sidereal modulation. This last effect produces a splitting of the signal power among five frequencies $f_{0}$, $f_{0} \pm f_{\oplus}$ and $f_{0} \pm 2f_{\oplus}$, where $f_{\oplus}$ is the Earth's sidereal frequency. 
In a more general formalism where a dual-harmonic emission is assumed, the strain at the detector has two components $h_{21}$ and $h_{22}$, representing the emission at $f_{\rm spin}$ and $2f_{\rm spin}$, respectively (see, e.g., Equations (1) and (2) in~\cite{LVKO1O2O3targeted2020}). The two harmonics are typically represented by the two amplitudes $C_{21}$ and $C_{22}$.  
Equation  (\ref{eq:detstrain}) is equivalent to the $h_{22}$ component only, i.e., which corresponds to the case of $C_{21}=0$ and $C_{22}=h_{0}/2$. 

When the signal is buried in the detector data, it is not possible to appreciate the signal morphology by eye since it is several orders of magnitude lower than the typical data strain amplitude. For this reason, effective data analysis methods are needed to extract the signal from the noise. Some of these methods are reported in the following section, while the results will be discussed in Section \ref{sec:results}.

\subsection{Search Methods}
\label{ssec:searchesmeth}
Given the long-lived and weak nature of CW signals, it is intuitive to think that the more data that are used in the analysis, the more signal power is accumulated, thus increasing the signal-to-noise ratio. The optimal search method to look for these signals, provided that the waveform is completely known and under the assumption of gaussian noise, is the matched filtering. Several methods have been developed to look for sources such as know pulsars, where all the parameters of the source are known. 
Source parameters may be available from EM observations, i.e., the source sky position and its rotational parameters are known. In this case, the so-called ``targeted'' searches are performed using matched filtering. Several implementations of the matched filtering techniques, also known as coherent searches, exist based on different detection statistics or algorithms: the \texttt{5-vector}~\cite{Astone5vec2010,Astone5vec2012} method, the \texttt{F/B/G statistic}~\cite{Jaranowski1998,Prix2009,Prix2007,JaranowskiFGstatistics2010,Cutler2005} and the time-domain heterodyne-based pipelines (\texttt{Bayesian} and \texttt{Band-Sampled-Data})~\cite{DupuisBayesian2005,pitkin2017,PiccinniBSD2018}. 

In targeted searches, the phase of the CWs is assumed to be locked to the rotational phase of the crust of the star, which is completely defined by its EM observation. When this assumption is relaxed, ``narrow-band'' searches are performed. This would allow taking into account, e.g., a differential rotation between the rigid crust and superfluid parts of the star.
In these searches, a small region around the EM-inferred signal parameters is investigated. In general, a narrow-band search for a given target is less sensitive than its respective targeted case due to the increased trial factor. 

As the knowledge about the source parameters decreases, the parameter space to investigate explodes, making  the use of fully coherent searches impractical~\cite{Brady1998}. For this reason, many semi-coherent techniques have been developed~\cite{Brady2000,Cutlershierarchical2005}. In a typical semi-coherent search, chunks of data may be first analyzed coherently (e.g., using a matched-filter) and then combined incoherently (i.e., not considering the phase information). This incoherent combination can be achieved in many different ways, for instance, by summing the detection statistics in each chunk of data analyzed and computing a single final detection statistic (see \cite{TenorioCW2021} for a detailed review of wide parameter space searches for CWs).
These searches are suitable in the case when no EM counterpart is present, such as all-sky search surveys for NS or DM candidates, or when the true nature of the source is not completely known, for instance, directed searches pointing to interesting sky regions or supernova remnants hosting potential young NS. 
Semi-coherent methods include the \texttt{FrequencyHough}~\cite{AstoneMethodFH2014,PalombaMethAFH2005,AntonucciMethodFH2008}, the \texttt{SkyHough}~\cite{KrishnanMethodSH2004,SintesMethodSH2006,JordanaSH2010}, the \texttt{Time-Domain $\mathcal{F}$-statistic}~\cite{Jaranowski1998,AasiFstatallsky2014,PisarskiMethodFstat2015}, \texttt{Weave}~\cite{WetteWeave2018}, \texttt{PowerFlux}~\cite{DergachevPowerFluxTech,LVKO3aallsky2021}, and 
\texttt{Einstein@Home}, based on the global correlation transform method~\cite{Pletsch2008,Pletsch2009,Pletsch2010}. A comparison of methods for the detection of GWs from unknown NSs is given in~\cite{Walsh2016}.

Loosely coherent methods, such as the \texttt{Falcon} pipeline~\cite{DergachevMethLooseCoh2010,DergachevMethLooseCoh2012,DergachevMethLooseCoh2019}, have been developed to manage the huge computing cost of all-sky searches while keeping the robustness to a wide family of signals, allowing for a loose phase track in the evolution of the signal frequency. 
A different approach can be used if one wants to go deep with the sensitivity in an all-sky search but keep the focus on a specific narrow frequency band, as described in~\cite{Wette2021}. 

\textls[-15]{Semi-coherent or incoherent methods can be used when no assumption on the source frequency evolution is made or if it is different from the standard Taylor expansion in time such as in Equation  (\ref{eq:phase}). These include the hidden Markov model \texttt{Viterbi} \mbox{tracker~\cite{SuvorovaMethodHMMNS2016,SunViterbidualharm2019,SunViterbiSNR2018},} the \texttt{Sidereal Filter}~\cite{DantonioMethSF2021} and cross-correlation based methods such as \texttt{CrossCorr}~\cite{RomanoCrossCorr2017} \texttt{STAMP}~\cite{ThraneSTAMP2011}, \texttt{SOAP}~\cite{BayleyMethodSOAP2019} or \texttt{Radiometer}~\cite{MitraRadiometer2008}. Some of these methods---CrossCorr method~\cite{WhelanMethodCrossCorr2015}, Viterbi/J-statistic method~\cite{SuvorovaMethodHMM2017}, the 5-vector binary method~\cite{Singhal5VECBIN2019} and the method in~\cite{Leaci2017}---have been adapted for the LMXB searches such as Sco-X1, where also the orbital modulation needs to be considered and a stochastic spin wandering, including spin-up, is sometimes taken into account. Methods specifically tailored for all-sky searches for unknown binaries are the BinarySkyHough~\cite{CovasMethodBinSH2019} and TwoSpect~\cite{GoetzTwoSpec2011}.}

Different flavors of the above-mentioned techniques have been adapted for the case of long-transient CW signals~\cite{OliverMethTrans2019,MillerMethodGFH2018,Prixtransients2011,Suntransient2019,BanagiriViterbiSTAMP2019,Keiteltransient2016,Keiteltransient2018} and glitching pulsars~\cite{Ashtonglitch2018,Ashtonglitch2017}, ensemble of pulsars~\cite{Cutler2005,PitkinEns2018,BuonoEns2021} or DM candidates~\cite{DantonioMethBC2018,IsiBosonmeth2019,MillerDPmeth2021}. Furthermore, several methods have been developed for the search of CW signals using machine learning~\cite{MorawskiML2019,YamamotoML2020,MillerML2019,BeheshtipourML2020,DreissigackerML2019,DreissigackerML2020,BayleyFU2020}.
For the latest review on CW methods and searches, see~\cite{RilesCW2017,SieniawskaCW2019,TenorioCW2021}.

In general, all these pipelines produce the so-called search candidates, i.e., potential astrophysical signals that need to be investigated in detail. These candidates undergo a series of post-processing veto steps, and then, the most interesting ones are followed up with different techniques in order to clearly establish their significance, e.g.,~\cite{LeaciFU2015,TenorioFU2021,TenorioFU22021,IntiniFU2020,ZhuFU2017,KeitelFU2016,AshtonFU2018,IsiFU2020,BayleyFU2020,BehnkeFU2015,KeitelFU2014,tenoriodistromax2021}.
A fraction of these candidates is vetoed if found near known instrumental artifacts, also known as spectral lines, typically present in the detectors~\cite{Covas2018}. Multiple approaches exist for this part of the analysis to discard or pass the outliers to the follow-up stage, as described in~\cite{TenorioCW2021}.
Sometimes the assessment of the real origin of an outlier is the most difficult and less automatized part of the analysis, but in general, for most of the follow-up techniques, the final goal is to check if the outlier significance follows an expected theoretical trend. This is achieved, for instance, by checking if the outlier signal-to-noise ratio grows as more data are used to analyze that candidate or by increasing the coherence time. Alternative follow-up methods are based on machine learning techniques (see Section 4 in~\cite{TenorioCW2021}).

If all the outliers are discarded from being real astrophysical signals during the follow-up stage, no CW signal detection can be claimed. In this case, typically upper limits on the signal strain are provided. These values usually span the full frequency range investigated in the search for the case of wide parameter space searches. For a given confidence level, the upper limit curve determines a watershed on the signal amplitude $h_{0}$ above which it is possible to exclude the existence of CW signals. These curves are estimated using a tremendous number of simulated signals added to the data or via less computationally expensive methods based on analytically derived or scaling formulas (the most famous being the "sensitivity depth" %MDPI: Is the italics necessary? NO
~\cite{Dreissigacker2018,BehnkeFU2015}, see also~\cite{DergachevULmeth2013,FeldmanCousins1998}). 
In addition to giving an estimate of how deep a given search can look for a CW signal, upper limits are often used to set astrophysical constraints on some related quantities according to the type of search performed. Typical constraints derived from upper limit curves are those on the ellipticity of the spinning NS (or the quadrupole $Q_{22}$) or the r-mode amplitude $\alpha$. In searches looking for DM candidates, these limits can be converted to, e.g., a coupling constant, mass exclusion regions, merging rates and abundances.

According to population studies, an order of $10^8$--$10^9$ 
NSs is expected to exist in our Galaxy~\cite{Treves2000}.  Only a small fraction of the expected population has been detected electromagnetically or has an associated EM counterpart~\cite{Rajwadepop2018}. Among all the \mbox{${\sim}3000$ sources} reported in the Australian Telescope National Facility (ATNF) catalog~\cite{ManchesterATNF2005}, about $10\%$ are in a binary system. If sources spinning at frequencies above 10 Hz are considered, this fraction increases to  ${\sim}50\%$.
Information provided by astronomers is used to constrain the investigated parameter space. In this sense, CW searches have a strong multi-messenger approach.  EM information is fundamental for the search of known pulsars, but it provides interesting inputs for the search of new sources, such as supernova remnants potentially hosting an NS, as well as updated information about these sources.
In particular,  the detection of CWs from a particularly interesting source or sky region (galactic center, globular clusters, etc\ldots ) represents a real discovery; given the long-lasting nature of the signal, EM observations following the detection could be carried out for a long time, opening the possibility to, e.g., perform tests of GR and/or to set interesting constraints on the inner composition of the emitting body and on its EOS.
On the other hand, the discovery of EM dark sources, including DM candidates, will provide a new tool for the study of a new population of astrophysical sources, as well as contribute to solving many of the open questions about our Universe. 

\section{Recent Results}
\label{sec:results}
Since the beginning of the advanced detector era, the interest with respect to CW sources has increased, as probed by the increasing number of publications on this topic over the years. Boosted by the enhancement of the sensitivity of the detectors as well as the development of data analysis techniques,  different and new searches have taken place. I briefly review the most recent search results for CW signatures in advanced detector data from the latest three observing runs~\cite{GWTC-1,GWTC-2,GWTC-3}.
The first observing run (O1) started on  11 September 2015 and finished on  19 January 2016. The second observing run (O2) lasted from  30 November 2016 to  25 August 2017. During O1, only the two LIGO interferometers were recording data, while during O2, the Advanced Virgo joined the run at the beginning of August 2017. The O3 run started on  1 April 2019, 15:00 UTC and finished on  27 March 2020, \mbox{17:00 UTC}. In the following, we refer to the period 1 April 2019--1 October 2019 as O3a, while the second part of O3, O3b, starts on 1 November 2019 after a month-long commissioning break. Wherever possible, the latest and most stringent results will be presented at the beginning of each section, while older results will follow.

\subsection{Results from Known Sources (Pulsars, LMXBs, Supernova Remnants)}
\label{ssec:resultsknown}
In this section, I review the latest results for searches of CW from known sources or sources with an EM counterpart, such as the central compact objects in supernova remnants. 

\subsubsection{Known Pulsars}
A targeted search is the most sensitive way to look for persistent signals from known pulsars. These searches strongly rely on the EM information available about the source. Indeed, in these searches, the GW  phase evolution is completely determined by the EM observations of the rotational parameters of the star. This in principle gives a full picture of the GW emission from the star.
A slightly different type of search, known as narrow-band, is sometimes applied if the coupling model between the GW and EM signal phase evolution is relaxed. This allows to assume a more generic scenario, and typically, a small region around the expected GW frequency and spin-down is investigated. 

\begin{figure}[H]
%\centering
\includegraphics[scale=0.5]{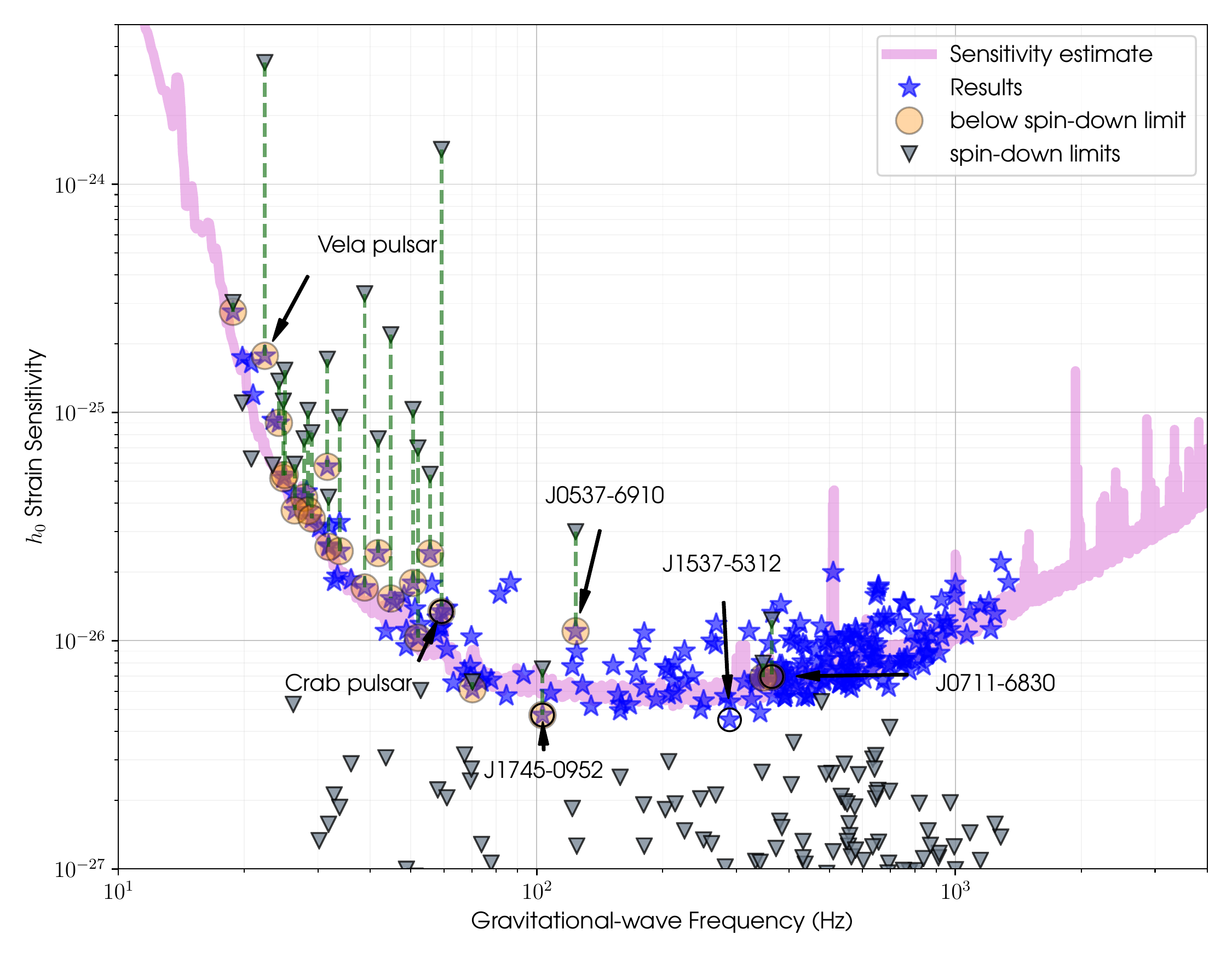}
\caption{Upper limits on $h_0$ for the 236 pulsars in the targeted search in~\cite{LVKtargetedO2O3236pulsar2021} 
 are relative to the time-domain Bayesian method. The stars show $95\%$ credible upper limits on the amplitudes of $h_0$. Grey triangles represent the spin-down limits for each pulsar computed as in Equation  (\ref{eq:spindownlimit}). Pulsars with upper limits surpassing their spin-down limits are marked with yellow circles. 
The pink curve gives an estimate of the expected strain sensitivity of all three detectors combined during the course of O3. Figure taken from~\cite{LVK	targetedO2O3236pulsar2021}}
\label{fig:knownpulsars}
\end{figure}

The latest extensive search for CW from known pulsars is reported in~\cite{LVKtargetedO2O3236pulsar2021}. In this work, 236 pulsars have been analyzed, among which the Crab and the Vela pulsars, using LIGO and Virgo O3 data combined with O2 data. Several pulsars glitched during the considered runs and some have sufficient information from EM observations on their orientation to restrict their priors on the $\iota$ and $\psi$ parameters. The ephemerides come from the observations from the CHIME, Hobart, Jodrell Bank, MeerKAT, Nancay, NICER and UTMOST observatories.
The search was run using three independent pipelines: the time-domain Bayesian method~\cite{DupuisBayesian2005,pitkin2017}, the F/G/D-statistic method~\cite{Jaranowski1998,JaranowskiFGstatistics2010,Verma2021} and the 5-vector method~\cite{Astone5vec2010,Astone5vec2012}. All the pipelines searched for the standard single harmonic GW emission from the $l = m = 2$ mass quadrupole mode with a frequency at twice the pulsar rotation frequency. In addition, the Bayesian and the F/G/D-statistic pipelines assumed a dual harmonic emission scenario, i.e., the $l = 2;m = 1,2$ modes with a GW frequency at once and twice the rotation frequency~\cite{Jones2010,PitkinDH2015}. To complete the search, the 5-vector method limited the search to the single harmonic scenario but at the $l = 2;m = 1$ mode only, hence assuming a GW frequency at the star rotation frequency. Furthermore, the D-statistic has been used to test the dipole radiation as predicted by the Brans--Dicke theory~\cite{Brans1961,Verma2021}. No GW detection has been reported, and upper limits on the strain are given in Figure \ref{fig:knownpulsars}. Blue stars indicate the upper limit of a given pulsar, and those with upper limits below the spin-down limit (see Equation  (\ref{eq:spindownlimit})) are marked with yellow circles. Grey triangles mark the spin-down limit of each pulsar. The pink curve is a sensitivity estimate for a generic targeted search, and it is given by ${\sim}10\sqrt{S_{n}(f)/T_{\rm obs}}$, where $S_{n}(f)$ is the power spectral density.

From Figure \ref{fig:knownpulsars}, it can be seen that 23 out of the 236 pulsars have strain amplitudes lower than the limits calculated from their electromagnetically measured spin-down rates. 
Among these, 90 millisecond pulsars have a spin-down ratio, i.e., the ratio between the upper limit and the spin-down limit, less than 10. For instance, J0437-4715 and J0711-6830 have spin-down ratios of 0.87 and 0.57, respectively. For the Crab and Vela pulsars, these limits are factors of ${\sim}100$ and ${\sim}20$ lower than their spin-down limits, respectively. Nine of these twenty-three pulsars beat their spin-down limits for the first time. 
The pulsar with the smallest upper limit on $h_{0}$ was J1745-0952 with $4.72 \times 10^{-27}$. The best $Q_{22}$ upper limit was achieved by J0711-6830 with $4.07 \times 10^{29} \mathrm{~kg} \mathrm{~m}^{2}$, equivalent to a limit on the ellipticity of $5.26 \times 10^{-9}$. This result is lower than the maximum values predicted for
a variety of NS equations of state~\cite{Gittins2021}.
Concerning the dual harmonic search, the most constraining upper limit for $C_{21}$ was $7.05 \times 10^{-27}$ for J2302+4442. The stringent $C_{22}$ upper limit was $2.05 \times 10^{-27}$ for J1537-5312.

A previous search using O1 plus O2 data on 222 pulsars is given in~\cite{LVKtargetedO1O2222pulsar2019}. In that search, the percentage of GW emission contributing to the spin-down luminosity for Crab and Vela pulsars was less than $0.017\%$ and $0.18\%$, respectively. Compared to the O2+O3 search~\cite{LVKtargetedO2O3236pulsar2021} described above, these numbers decrease as expected to $0.009\%$ for Crab but increased for Vela up to a maximum of $0.27\%$. This unexpected result is due to the presence of a significant noise line in the LIGO Hanford detector at twice Vela's rotational frequency during O3. 
The O1+O2 search in~\cite{LVKtargetedO1O2222pulsar2019} and the O2+O3 
 search in~\cite{LVKtargetedO2O3236pulsar2021} certainly improve the results of the previous ``CW pulsar catalog'' for 200 pulsars in O1 data~\cite{LVKtargetedO1200pulsar2017}.

Tests for non-GR polarizations are also provided for the 23 pulsars beating the spin-down limit in~\cite{LVKtargetedO2O3236pulsar2021}, checking for the additional non-negligible scalar radiation originated from the time-dependent dipole moment~\cite{Verma2021}. No deviation from GR has been reported, and the stringent upper limit for dipole radiation is obtained for the millisecond pulsar J0437-4715. A previous test on GR from known pulsar searches, not assuming any particular alternative theory of gravity, is given in~\cite{LVCtestGRknownpulsars2018} using O1 data.

%\textcolor{red}{( Please ensure that permission has been obtained and there is no copyright issue. If copyright is needed, please provide a citation in the following format: “Reprinted/adapted with permission from Ref. [XX]. Copyright year, copyright owner’s name”. Same as other figures with citations. Please kindly check all.)}. Results %MDPI:

A separate dual harmonic search has been performed on the same dataset (O2+O3) for the energetic young pulsar and frequent glitcher PSRJ0537-6910 in~\cite{LVKtargetedO2O3J05372021}. No CW signal has been detected in this search but, for the first time, the spin-down limit for this type of GW emission on this source has been surpassed. Results show that GWs from the $l = m = 2$ mode contribute to less than 14\% of the spin-down energy budget. The same target has been investigated by two other searches, tuned for r-modes oscillations, using ephemerides from NICER~\cite{Ho2020NICER} in  O3~\cite{LVKtargetedO3rmodeJ05372021}  and  O1+O2~\cite{FesikrmodeO1O22020} data. As discussed in Section~\ref{ssec:NStheo}, the relation between the gravitational-wave frequency and the star spin frequency in r-mode emission scenarios strongly depends on the NS structure. For this reason, both searches assumed the frequency evolution model proposed in~\cite{Caride2019}.
Searches for the r-mode emission from PSRJ0537-6910 are motivated by the fact that this pulsar shows  a trend in the inter-glitch behavior that suggests an effective braking index close to $n=7$ at long times after the glitch~\cite{AnderssonJ05372018,AntonopoulouJ05372017}. A braking index of seven is expected for a spin-down dominated by the r-mode emission.
The amplitude's upper limits in the O3 search~\cite{LVKtargetedO3rmodeJ05372021} improve the existing results by a factor of up to three, placing stringent constraints on theoretical models for r-mode-driven spin-down in PSR J0537-6910. 
Another search for r-mode emissions for the Crab pulsar is reported in~\cite{RajbhandarirmodeO1O22021}, providing for the first time upper limits beating the spin-down limit for this source/type of emission in O1 plus O2 data.

Five sources have been the target of a previous single and dual harmonic search for known pulsars~\cite{LVKO1O2O3targeted2020}. These pulsars have been analyzed in the data from the first two and the first half of the third observing run of advanced detectors (O1+O2+O3a). The targets included three recycled pulsars\endnote{Recycled pulsars are ordinary pulsars that have been spun up by accretion from a companion star in a binary system.} (J0437-715, J0711-6830, and J0737-3039A) and the two young pulsars Crab and Vela. For Crab, Vela and J0711-6830, a narrow-band search has also been performed using the 5-vector narrow-band method~\cite{MastrogiovanniNB5V2017,AstoneNB5V2014}. Results from this work constrain the equatorial ellipticities of some of the targets to be less than $10^{-8}$, and for the first time, an upper limit below the spin-down limit for a recycled pulsar is reported. This latter aspect is interesting since the evolutionary history of a recycled pulsar is connected to an early accretion phase from a binary companion, very different from the case of young and more slowly spinning pulsars. 
Another set of seven recycled pulsars is investigated in~\cite{AshokAreciboO1O2O3a2021}. The timing information is provided by the  Arecibo 327 MHz Drift-Scan Pulsar Survey, and the search is performed on O1+O2+O3a data. All the targets, except for the millisecond pulsar J0154+1833, are in binary systems. 
The stringent constraint on the ellipticity is $1.5 \times 10^{-8}$ for PSR J0154+1833.

Data from Fermi-LAT observations have been used in the search for CW from single harmonic emission from PSR J0952-0607~\cite{NiedertargetedbinaryO1O2J0952_06072019} and dual-harmonic emission from PSR J1653-0158~\cite{NiedertargetedbinaryO1O2J1653-01582020}, using data from O1 and O2.
The results of the search for CW signals from ten sources associated with TeV sources observed by the High Energy Stereoscopic System (H.E.S.S.)  are reported in~\cite{BeniwalHess2021}. All the sources are associated with isolated NSs and with estimated distances up to 7 kpc. In particular, two targets are $\gamma$-ray pulsars, while eight H.E.S.S. sources are pulsar wind nebulae powered by known pulsars.
The ephemerids for each pulsar are provided by the ATNF pulsar catalog. The search is performed using O2 data and tracks, with a hidden Markov model, the three GW frequencies expected to be at 1x, 2x and 4/3x the source spin frequency in a narrow band of 1 Hz around \mbox{each multiple.} %Please confirm intended meaning is retained. -> OK

The latest results from the O3 fully coherent narrow-band search can be found in~\cite{LVKnarrowbandO32021}. The search looks for CW from 18 pulsars using the five-vector and F-statistic narrow-band pipelines. For seven of these pulsars, the upper limits are lower than their spin-down limit. These results overcome the corresponding ones from the narrow-band search in O3a~\cite{LVKO1O2O3targeted2020}, improving the upper limits by a factor ${\sim}35\%$ for the Crab pulsar, and ${\sim}10\%$ for Vela and J0711-6830 pulsars even if the parameter space investigated in this search is 6.5, 1.2 and \mbox{7.1 times} larger.
Results from an additional search for long-duration transient GWs after pulsar glitches are also reported in~\cite{LVKnarrowbandO32021}. Six targets have been investigated by the long-transient search, performed using the transient F-statistic method~\cite{modafferitransientsetup2022,Prixtransients2011}. For these six sources, showing a total of nine glitches, an independent long-transient search is performed after each glitch.
None of these targets, neither from the narrow-band or long-transient search, produced an outlier significant enough to be considered as a signal, although two marginal outliers are reported for the last glitch of 
J0537-6910.
The transient F-statistic method has been applied for the search of long-duration transients after glitches in the Vela and Crab pulsars in O2 data~\cite{KeiteltransientO22019}. Previous narrow-band searches using O2 and O1 data are given in~\cite{LVKnarrowbandO22019,LVKnarrowbandO12017}.

%_________________________________________

\subsubsection{Supernova Remnants}
Another family of interesting targets with EM counterparts, but without precise timing of the rotational parameters, is the supernova remnants. Supernova remnants may host young NSs, which are excellent targets for the search for CWs. Given the lack of information on the spin frequency of the central compact object inside the nebula, directed searches are performed. In these searches, the only information assumed as known is the source sky position, while the remaining parameters $(f,\dot{f},\ldots )$ are investigated in a wide range. The choice of the parameter space to investigate is only limited by the computing cost of the search, and this may change according to the different algorithms used and to the specific setup chosen. Given an amount of computing power, some regions may have a higher priority over the full GW sensitivity band~\cite{Ming2016,Ming2018}. No CW detection is reported from these sources. In the following, I report the latest search results from these targets.

\begin{figure}[H]
%\centering
\includegraphics[scale=0.32]{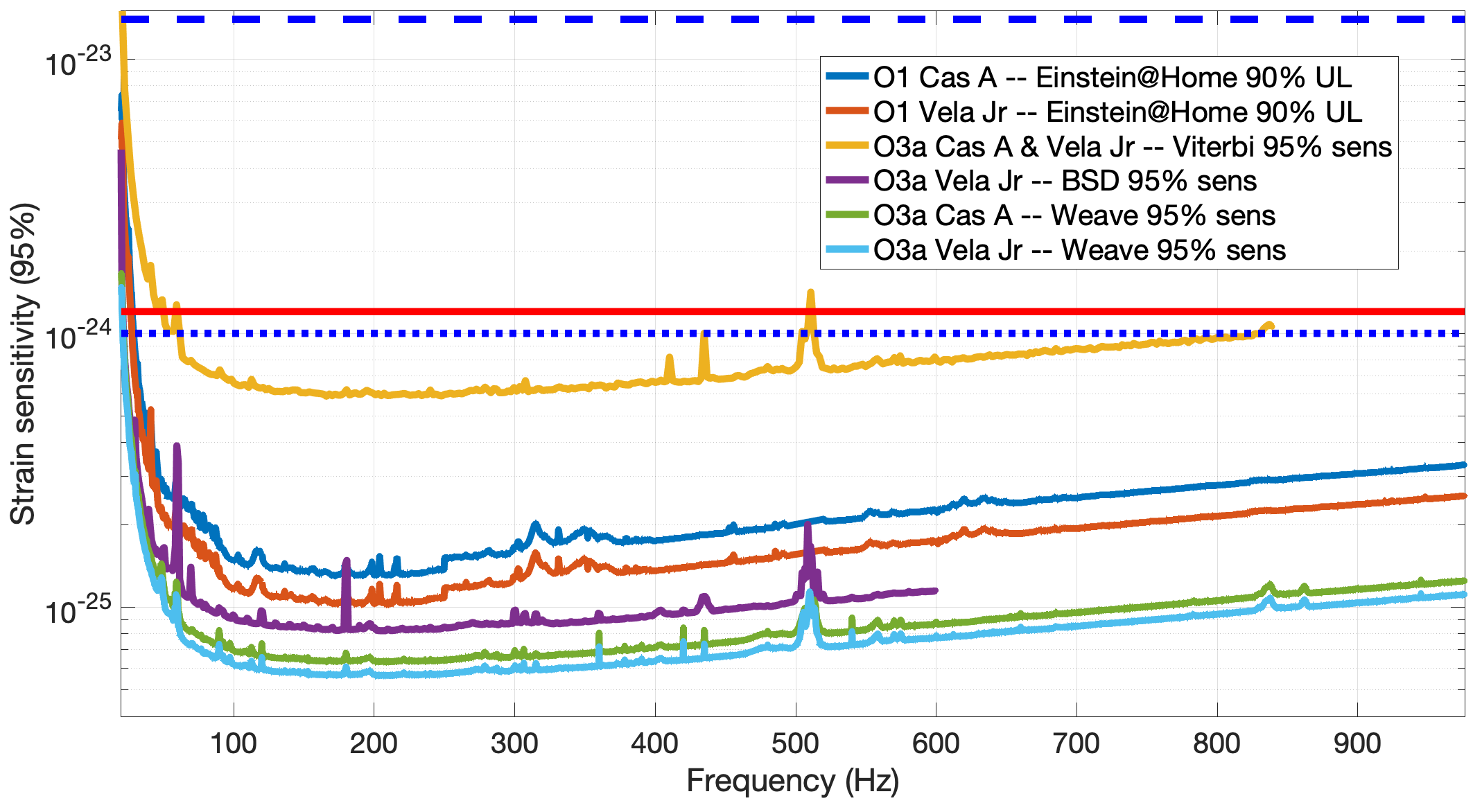}
\caption{Results %MDPI: Please change hyphen into minus sign in figure ->can'tt
taken from the supernova remnant search in LIGO O3a data for Cas A and Vela Jr using Weave~\cite{LVKSNRCasAVelaJrO3aWeave2021}.  Estimated GW strain amplitude sensitivities are given at the $95\%$ confidence level. Additional results from prior searches for Cas A and Vela Jr. in O1 Einstein@Home $90\%$ C.L.~\cite{MingSNREatHO12019} and the O3a model-robust Viterbi method (yellow) and band-sampled-data method (purple)~\cite{LVKSNRBSDviterbisO3a2021}.
The solid red horizontal line indicates the age-based upper limit on the Cas A strain amplitude. The dashed (dotted) horizontal blue lines indicate the optimistic (pessimistic) age-based upper limit on Vela Jr. strain amplitude, assuming an age and distance of 700 yr and 0.2 kpc (5100 yr and 1.0 kpc).  Figure adapted from~\cite{LVKSNRCasAVelaJrO3aWeave2021}.} %MDPI: Please ensure that permission has been obtained and there is no copyright issue. If copyright is needed, please provide a citation in the following format: “Reprinted/adapted with permission from Ref. [XX]. Copyright year, copyright owner’s name”. More details on “Copyright and Licensing” are available via the following link: https://www.mdpi.com/ethics#10.
\label{fig:directedSNR_casAVelaO3a}
\end{figure}

Searches for supernova remnants in LIGO-Virgo O3a data are reported in~\cite{LVKSNRCasAVelaJrO3aWeave2021,LVKSNRBSDviterbisO3a2021}. To date, the results in~\cite{LVKSNRCasAVelaJrO3aWeave2021} are the most stringent for the source targeted: Cassiopeia A (Cas A, G111.7-2.1) and Vela Jr. (G266.2-1.2) supernova remnants. Both sources are extremely young, with inferred lower estimate ages of ${\sim}300$ years for Cas A, and ${\sim}700$ years for Vela Jr. With these assumptions, these targets have a corresponding indirect age-based limit (see Equation  (\ref{eq:agebasedlimit})) of ${\sim}1.2 \times 10^{-24}$ for the central compact object in Cas A, assuming a distance of 3.3 kpc, and $1.4\times 10^{-23}$ for Vela Jr. if located at 0.2 kpc.  The search is based on the semi-coherent Weave method~\cite{WetteWeave2018}. No GW signal is detected in the analyzed band, and the best sensitivity to the signal strain amplitude reached in this search is ${\sim}6.3 \times 10^{-26}$ for Cas A and ${\sim}5.6 \times 10^{-26}$ for Vela Jr., well below their age-based upper limit. 
A comparative plot is shown in Figure \ref{fig:directedSNR_casAVelaO3a}, showing  the upper limits provided by the searches in~\cite{LVKSNRBSDviterbisO3a2021,LVKSNRCasAVelaJrO3aWeave2021,PapaSNREatHO1O22020}. In the search reported in~\cite{LVKSNRBSDviterbisO3a2021}, three complementary pipelines have been used: the Band-Sampled-Data directed pipeline, the single harmonic Viterbi and the dual harmonic Viterbi pipelines. Each pipeline used different signal models and methods for the identification of noise artifacts. In particular, the two Viterbi pipelines are robust to sources with spin wandering. This search targets a total of 15 supernova remnants, including Cas A, Vela Jr. and G347.3-0.5. These three targets have also been analyzed  in a previous Einstein@Home search~\cite{MingSNREatHO12019} using O1 data. The most interesting candidates of this search have been followed up in O2 data, along with new sub-threshold candidates of the same O1 search as reported in~\cite{PapaSNREatHO1O22020}. A full deep search for CWs from G347.3-0.5 in the frequency band below 400 Hz in O2 LIGO data is described  in~\cite{MingSNREatHO22022}. This choice is justified by the existence of a sub-threshold candidate at around 369 Hz that survived  the search in~\cite{PapaSNREatHO1O22020}. The same candidate has been further investigated with an MCMC method in O2 data~\cite{TenorioFU2021}.
Any significant result has been reported from this search in O2 data, and hence, upper limits are provided, improving those from the same search in O1 data and those from the more robust single harmonic Viterbi tracker in O3a~\cite{LVKSNRBSDviterbisO3a2021}.

For the closest target Vela Jr., the best upper limit in~\cite{LVKSNRBSDviterbisO3a2021} is $8.2\times10^{-26}$ for the band-sampled-data pipeline. This result is more constraining than those set by~\cite{PapaSNREatHO1O22020,MingSNREatHO12019}; however, it is less stringent than the results in~\cite{LVKSNRCasAVelaJrO3aWeave2021} by a factor of $30\%$ (see Figure \ref{fig:directedSNR_casAVelaO3a}).
It should be noted that the searches in~\cite{MingSNREatHO12019,PapaSNREatHO1O22020} used O1 and O2 data, while the searches in~\cite{LVKSNRBSDviterbisO3a2021,LVKSNRCasAVelaJrO3aWeave2021} used data from detectors with improved sensitivity (O3a). However, the Einstein@Home O1 searches~\cite{MingSNREatHO12019,PapaSNREatHO1O22020} used a coherent integration duration ${\sim}$20--150 times longer than the one in the band-sampled-data search in~\cite{LVKSNRBSDviterbisO3a2021}, while the Weave search in~\cite{LVKSNRCasAVelaJrO3aWeave2021} used a coherence length 10--70 times longer. This guarantees a starting theoretical sensitivity gain of a factor of 2.1--3.5 and 1.7--2.9  for~\cite{MingSNREatHO12019,PapaSNREatHO1O22020} and ~\cite{LVKSNRCasAVelaJrO3aWeave2021}, respectively\endnote{The ranges are due to fact that the coherence time used in~\cite{LVKSNRBSDviterbisO3a2021} changes in each 10Hz frequency band.}. On the other hand, the use of longer coherence times corresponds to an increase in the computing power needed for the search, as it scales at least as the fourth power of the coherence time. Furthermore,  the use of a longer coherence time may negatively impact   the robustness of the search.
This makes it clear that different approaches are needed when wide parameter space searches are performed, choosing the right balance between computing cost, robustness and sensitivity.

Previous searches for CW from some of the 15 targets investigated in~\cite{LVKSNRBSDviterbisO3a2021} are presented in~\cite{LindblomSNRO22020,MillhouseSNRO22020} using LIGO O2 data, although using, respectively, a fully coherent method and a less sensitive but model robust Viterbi tracker. A previous fully coherent O1 search is given in~\cite{LVKSNRO1coh2019}, also listing among the targets  the extrasolar planet candidate Fomalhaut b suggested to be a nearby old
NS. The latter is also investigated  in~\cite{JonesSNRFomalhaut2021} using a hidden Markov model Viterbi tracker.

%________________________________
\subsubsection{Low-Mass X-ray Binaries and Sco-X1}
A significant fraction of observed known NS is in a binary system. 
Among these, LMXBs are of particular interest for CW searches. Indeed, the presence of a GW torque could explain why the maximum observed spinning frequency 716 Hz for PSR J1748-2446ad is well below the centrifugal breakup frequency, estimated at ${\sim}1400$ Hz.
These systems are typically formed by an NS (or stellar-mass BH) and a low mass (${\leq}1 M_{\odot}$) stellar companion.
A recent search for CWs from 20 accreting millisecond X-ray pulsars in O3 data is given in~\cite{LVKO3LMXB2022}. Five of these twenty accreting pulsars have  already been analyzed with the same method in O2 data~\cite{MiddletonO2HMMLMXB2020}.
The search relies on the EM observation of the spin frequencies and orbital parameters of the system, happening during the active outburst phase.  
The algorithm uses a hidden Markov model (implemented via Viterbi) to track spin wandering~\cite{SuvorovaMethodHMMNS2016} and combines its results with the J -statistic to account for the orbital phase modulation~\cite{SuvorovaMethodHMM2017}. For each target, the search looks for CW signals in sub-bands around $\{1,4 / 3,2\} f_{\rm spin}$. 
One of the targets, SAX J1808.4-3658, went into outburst during O3a. Given that, according to some models, CWs are only emitted during this outburst phase~\cite{HaskellJ10232017}, an opportunistic search for SAX J1808.4-3658 is also provided in~\cite{LVKO3LMXB2022}.
No significant candidates are reported from this search.  However, upper limits have been placed on the GW strain amplitude with $95 \%$ confidence level. The strictest constraint is  $4.7 \times 10^{-26}$ from IGR J17062-6143. The stringent constraints on the ellipticity for the $f_{\rm gw}=2f_{\rm spin}$ case is $\varepsilon^{95 \%}=3.1 \times 10^{-7}$ for the IGR J00291+5934 source.  The corresponding upper limit on the r-mode amplitude ($f_{\rm gw}=\frac{4}{3}f_{\rm spin}$) is $\alpha^{95 \%}=1.8 \times 10^{-5}$ for the \mbox{same source.}

\begin{figure}[H]
%\centering
\includegraphics[scale=5.5]{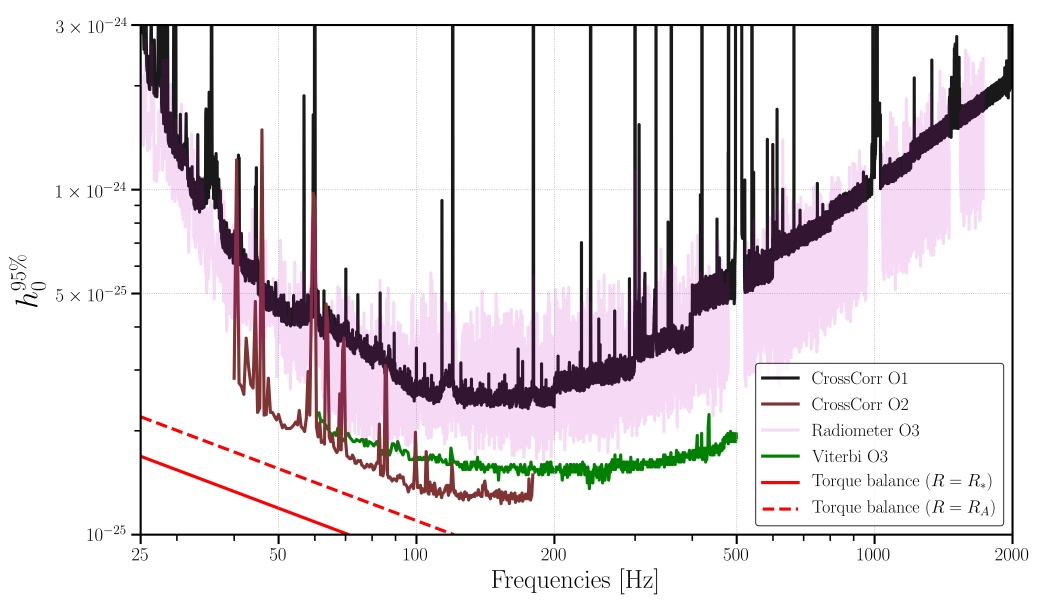}
\caption{GW strain upper limits at $95\%$ confidence as a function of frequency. No assumption on the $\iota$ angle is made for the curves in the figure.  The curves are for the CrossCorr O1 search~\cite{LVKO1CrossCorrScoX12017}~(black line), the CrossCorr O2 search~\cite{ZhangO2ScoX12021} (brown line), the Radiometer O3 search~\cite{LVKO1O3StochAnis2021} (light pink line), and the O3 hidden Markov model search~\cite{LVKO3HMMScoX12022} (green line). The indirect torque-balance upper limits (see Equation  (\ref{eq:torquebalancelimit})), for the $r_{m}=R$ case (red solid line) and for $r_{m}$ equal to the Alfven radius (dashed red line), are also plotted. Figure taken from~\cite{LVKO3HMMScoX12022}. }%MDPI: Please ensure that permission has been obtained and there is no copyright issue. If copyright is needed, please provide a citation in the following format: “Reprinted/adapted with permission from Ref. [XX]. Copyright year, copyright owner’s name”. More details on “Copyright and Licensing” are available via the following link: https://www.mdpi.com/ethics#10.
\label{fig:ScoX1}
\end{figure}

To date, several searches have targeted Sco X-1~\cite{ZhangO2ScoX12021,LVKO1HMMScoX12017,LVKO2HMMScoX12019,LVKO1CrossCorrScoX12017} in advanced detector data. The latest results using O3 data are reported in~\cite{LVKO3HMMScoX12022}. Sco-X1 is a very interesting candidate for CW searches since it is the brightest extrasolar X-ray source in the sky, which means that it is potentially an optimal source of GWs~\cite{Bildsten1998,Wagoner1984}. The system is composed of a $1.4~ M_{\odot}$ NS and a $0.7~M_{\odot}$ companion star~\cite{WangScoX2018}.
Although being a known source, no EM measurement of the rotation frequency and frequency derivative exists for Sco-X1. In addition to these uncertainties, some of the binary parameters are barely known, making the search for this source very computationally challenging. To date, no CW is reported for this source, while the torque-balance limit (see Equation  (\ref{eq:torquebalancelimit})) has been beaten for the first time in~\cite{ZhangO2ScoX12021}, using CrossCorr method~\cite{WhelanMethodCrossCorr2015,Meadors2018} in O2 data. The best constraints are as low as $7.06 \times 10^{-26}$, assuming an NS spin inclination $\iota=44^{\circ} \pm 6^{\circ}$~\cite{FomalontScoX12001,WangScoX2018}.
To date, none of the searches beats the limit if an isotropic prior on $\iota$ is made, as reported in \mbox{Figure \ref{fig:ScoX1}}. 
A different method has been used in the latest search for Sco-X1 in O3~\cite{LVKO3HMMScoX12022}, using a hidden Markov model pipeline implemented with the Viterbi algorithm~\cite{SuvorovaMethodHMM2017}. This method, although being less sensitive than the cross-correlation search,  provides a complementary tool to CrossCorr results.
The upper limits from the O3 search are on average ${\sim}3$ times lower than those from the O2 hidden Markov model search~\cite{LVKO2HMMScoX12019} and 13 lower than the same search in O1 data~\cite{LVKO1HMMScoX12017}. Previous searches using O1 data and the hidden Markov model and CrossCorr methods are reported in~\cite{LVKO1HMMScoX12017,LVKO1CrossCorrScoX12017}.
In O3, the lowest $h_{0}^{95 \%}$ for a hidden Markov model tracked signal are $4.56 \times 10^{-26}, 6.16 \times 10^{-26}$, and $9.41 \times 10^{-26}$ for circular ($\iota=0^{\circ}$), electromagnetically restricted ($\iota=44^{\circ}$) and unknown polarizations, respectively. Furthermore, for the hidden Markov model search, the torque balance limit is beaten for a fixed $\iota$ below $200~\rm Hz$ and for $r_{m}$ equal to the NS radius or the Alfven radius~\cite{LVKO1HMMScoX12017}.
By beating the torque balance limit for $r_{m}$ at the Alfven radius, it is possible to identify some exclusion region over the mass-radius-magnetic field combination for some equation-of-state models (see~\cite{ZhangO2ScoX12021} for a discussion). Constraints in the mass--radius plane are also possible when $r_{m}$ equals the NS radius.  \vspace{-6pt}

\subsection{Results from Unknown Sources (All-Sky, Spotlight Surveys, Dark Matter Candidates)}
\label{ssec:resultsunknown}
In the absence of EM-driven information, wide parameter space searches for unknown sources allow scouring the whole sky in search for EM silent sources.  These searches include all-sky surveys for isolated NS and NS in binary systems. It is also usual to look for CWs from a specific and over-populated region in the sky, such as the Galactic Center or some globular clusters. In these cases, sometimes the information provided by EM observations is not enough to shed light on the exact population of the analyzed region, and GWs could help in this direction.
Finally, through GWs, we could also look for DM candidates.

\subsubsection{All-Sky Surveys}
\begin{figure}[H]
%\centering
\includegraphics[scale=0.5]{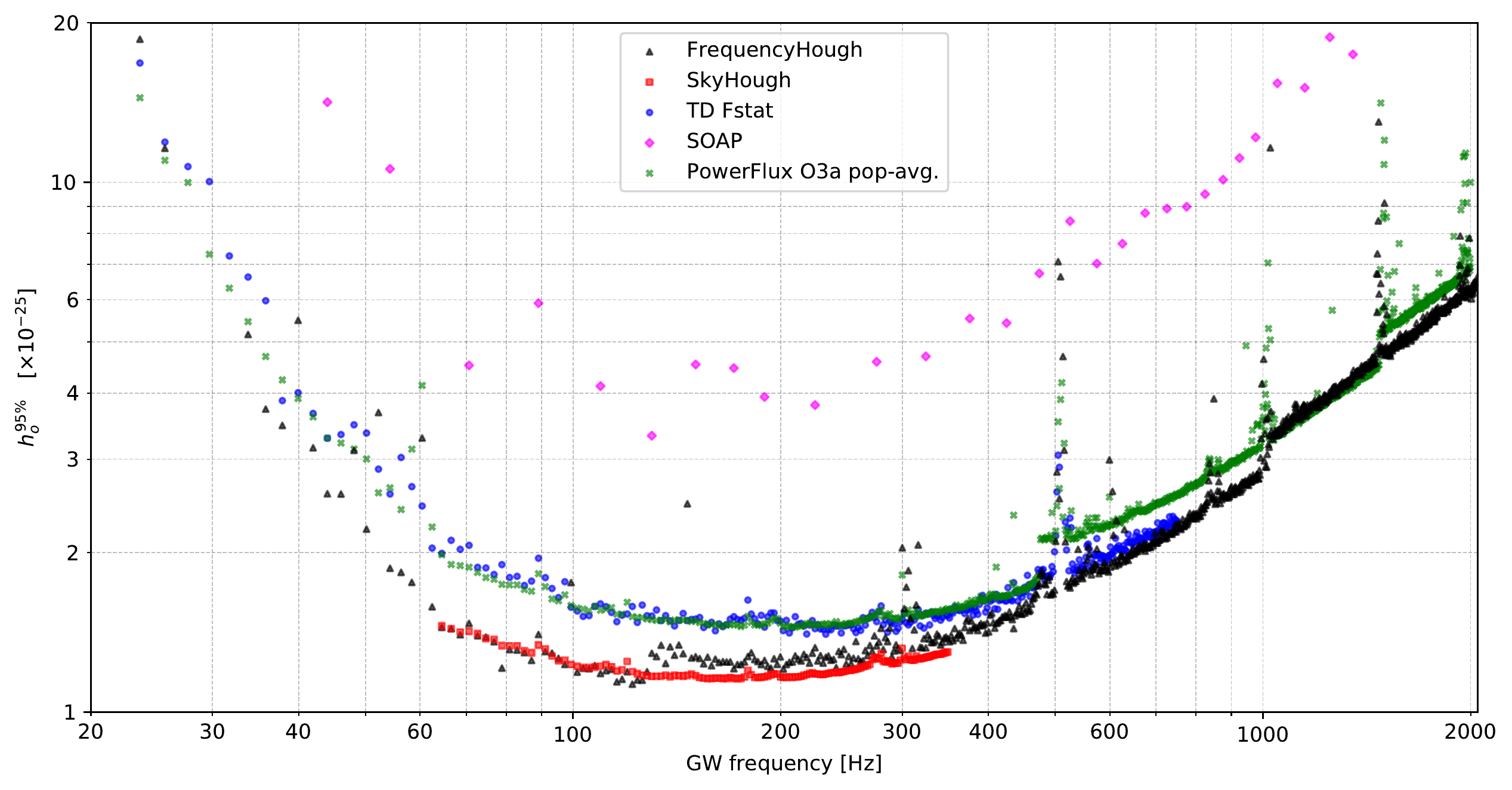}
\caption{Comparison of 95\% confidence upper limits on GW amplitude obtained in the O3 all-sky search by the FrequencyHough pipeline (black triangles), the SkyHough pipeline (red squares), the Time-Domain F-statistic pipeline (blue circles), and the SOAP pipeline (magenta diamonds). Population-averaged upper limits obtained in the Powerflux O3a search are marked with dark-green crosses. Figure taken from~\cite{LVKO3aallsky2021}.}
\label{fig:all_sky_isolatedO3}
\end{figure}

The latest results for all-sky surveys using LIGO-Virgo O3 data are reported in~\cite{LVKO3allsky2022} and a summary of the results can be found in Figure~\ref{fig:all_sky_isolatedO3}. Four different analysis methods are used to look for signals from isolated non-axisymmetric NSs: the FrequencyHough~\cite{AstoneMethodFH2014}, Sky-Hough~\cite{KrishnanMethodSH2004}, Time-Domain F-statistic~\cite{Jaranowski1998,AasiFstatallsky2014}, and SOAP~\cite{BayleyMethodSOAP2019}, and for the first time, applied in an all-sky CW search. 
The frequency investigated ranges from 10 to 2048 Hz and from $-10^{-8}$ to $10^{-9}~\rm Hz/s$ for the first frequency derivative\endnote{except for the SOAP pipeline that covers the additional small region $[1000;2048]~\rm Hz$ in frequency and $[10^{-9};10^{-8}]~\rm Hz/s$ in spin-up.}, each pipeline covers a subset of this parameter space except for the FrequencyHough. The frequency range is wide enough to cover most of the expected sources: young and energetic NS, millisecond pulsars and signals from r-mode oscillations. The search results are generic and are valid for any quasi-monochromatic, persistent signal characterized by the parameters investigated in this search following the linear phase evolution in Equation ~(\ref{eq:phase}). Constraints on the rate and abundance of inspiraling planetary-mass and asteroid-mass PBHs are  also discussed in this search.
No GW observation has been reported, and the best constraint on the strain amplitude at the 95\% confidence level is ${\sim}1.1 \times 10^{-25}$ in the 100--200 Hz frequency range. This is the first all-sky search using Advanced Virgo data.  

%MDPI: Please ensure that permission has been obtained and there is no copyright issue. If copyright is needed, please provide a citation in the following format: “Reprinted/adapted with permission from Ref. [XX]. Copyright year, copyright owner’s name”. More details on “Copyright and Licensing” are available via the following link: https://www.mdpi.com/ethics#10.

A previous search using the PowerFlux pipeline on O3a data~\cite{LVKO3aallsky2021} provided an upper limit on the strain, for the most favorable orientation, of $6.3 \times 10^{-26}$, improving previous results from O2 data. If compared with the population-averaged upper limit, the lowest $95\%$ C.L. in O3a is $1.4 \times 10^{-25}$, less constraining than the full O3 search in~\cite{LVKO3allsky2022} (see Figure~\ref{fig:all_sky_isolatedO3}). Results from this search have been used to constrain the rate and abundance of inspiraling PBHs in~\cite{millerPBHO3aPowerflux2021}.
To date, only the O2 Falcon search beats the upper limits in~\cite{LVKO3allsky2022}, although the frequency derivative range is at least $10^3$ smaller, making the results in~\cite{LVKO3allsky2022}, de facto, the most stringent limits for sources with spin-down as low as $-1\times 10^{-8}~\rm Hz/s$ in the 20--2000~$\rm Hz$ frequency range.

Upper limits on the ellipticity are also derived in~\cite{LVKO3allsky2022}. The search is able to detect nearby sources, within 100 pc, with ellipticities above $3\times10^{-7}$ emitting at 200 Hz, or even ellipticities as low as ${\sim}2\times10^{-9}$ for sources spinning at the highest frequencies.  For sources located as far as the Galactic Center region (8--10~kpc), the search is able to detect signals with ellipticities bigger than ${\sim}3\times10^{-6}$ for frequencies above 1500 Hz. 
Probing very small ellipticities with such a wide parameter space search certainly goes in the right direction for a detection. Indeed, ellipticities below $10^{-7}$ are in the expected range of sustainable strains in NS crusts~\cite{Gittins2020,Gittins2021,Ushomirsky2002,Haskell2006,Johnson-McDaniel2013}.
Previous searches performed using O2 data are given in~\cite{DergachevFalconO2allskyB2020,DergachevFalconO2allskyC2021,DergachevFalconO2allskyA2021,SteltnerO2EatHallsky2021,LVKO2allsky2019}, and O1 searches can be found in~\cite{DergachevO1Falconallsky2020,DergachevO1Falconallsky2019,LVKO1allskyE@H2017}.

No CW detection is reported  from all-sky searches for unknown NSs in binary systems. The latest search, carried out on O3a data~\cite{LVKO3aallskybinary2021,tenoriomoriondbinaryallskyO3a2021}, placed upper limits on the strain at $2.4 \times 10^{-25}$ for NSs in binary systems spanning orbital periods of 3--45 days. The search, carried out using the BinarySkyHough pipeline in O3a, is the update of the O2 all-sky binary survey presented in~\cite{CovasO2allskybinary2020}, reporting an improved result by a factor of 1.6. A good fraction of known NS is in a binary system. Despite this, no other all-sky search, specifically tailored for unknown binaries, has been carried out in advanced detector data. 
The main reason for this lack is due to the huge computing cost associated with these searches, significantly overcoming the already computationally demanding all-sky searches for isolated NS. However, according to~\cite{Singh2019}, for some combinations of the binary period and semi-major axis parameters, results from wide parameter space searches for unknown isolated sources are also valid for the binary case.

\subsubsection{Spotlight Surveys: The Galactic Center and Terzan 5}
Directed searches targeting interesting sky regions such as the Galactic Center and  the globular cluster Terzan 5 have been carried out in advanced detector data. Indeed, concerning the Galactic Center,  a significant population of up to hundreds or even thousands of NSs is expected to exist~\cite{Rajwade2017,Kim2018}.   
Moreover, the true composition of this region, showing an extended gamma-ray emission from its center~\cite{FermiGCE2016,FermiGCE2021,HESS2016}, is still under debate and could be explained, for instance, by the presence of an unresolved population of millisecond pulsars~\cite{Bartels_2016, Calore2016,Gregoire2013,Hooper2018,Buschmann2020}. 
A recent search for the CW emission from the Galactic center region has been carried out in O3 LIGO and Virgo data~\cite{O3LVKGC}. The search uses a semi-coherent method developed in the band-sampled data framework and span a wider parameter space compared to the previous O2 search in~\cite{PiccinniGCO22020}. The frequency investigated ranges from $10~\rm Hz$ to $2000~\rm Hz$, while the spin-down/up lies between $[-1.8 \times 10^{-8}; 10^{-10}]\rm~Hz/s$.
%To date, the latest search for CW from the Galactic Center looks for signals in the frequency range $[10;710]~\rm Hz$ in O2 data~\cite{PiccinniGCO22020}. The search looks for CW signals from NS located within 25-150 parsecs from the Galactic Center. The spin-down range covered in this search range from $-1.8 \times 10^{-9}$ Hz/s to $3.7\times10^{-11}$ Hz/s. The search uses a semi-coherent method developed in the Band-Sampled-Data framework. 
No CW signal has been detected, and upper limits on the GW amplitude are presented. To date, this search reports the most stringent upper limit for targets located within \mbox{30--300 parsecs} from the Galactic Center, with a minimum strain of ${\sim}7.6 \times 10^{-26}$ at 142 Hz at a $95\%$ confidence level. The same search reports upper limits on the ellipticity and the r-mode amplitude. The same search has been used to identify exclusion regions in the BH/boson cloud mass plane for sources located in the Galactic Center.

A previous search for CW from the Galactic Center and Terzan 5 is reported in~\cite{DergachevGCT5O1} using O1 data. The search, based on a loosely coherent method, looks for signals emitting at frequencies in the range $[475;1500]~\rm Hz$  and with frequency time derivatives in the range $[-3.0;0.1]\times 10^{-8}$ Hz/s.
O3 data have  already been used to make a lower sensitivity search for an unresolved GW emission from the Galactic Center region using stochastic GW search methods~\cite{LVKO1O3StochAnis2021}. %An older search for this region has been performed in the fifth LIGO scientific run~\cite{LVKGCS52013}.

\subsubsection{Dark Matter Candidates---Ultralight Bosons and CDOs} 
Methods for the search of CW from NSs can be easily adapted to any kind of GW signal with features similar to the standard NS case. An emerging family of searches is the one for DM candidates.
This branch of investigations through CW tools is very immature when compared, e.g., with known pulsars searches.  Nevertheless, some searches have been carried out in advanced detector data. 
In Section \ref{ssec:DMtheory}, the major DM systems investigated in CW searches have been described: boson clouds around spinning BHs, ultralight vector bosons and CDOs.

\begin{figure}[H]
%\centering
\includegraphics[scale=0.25]{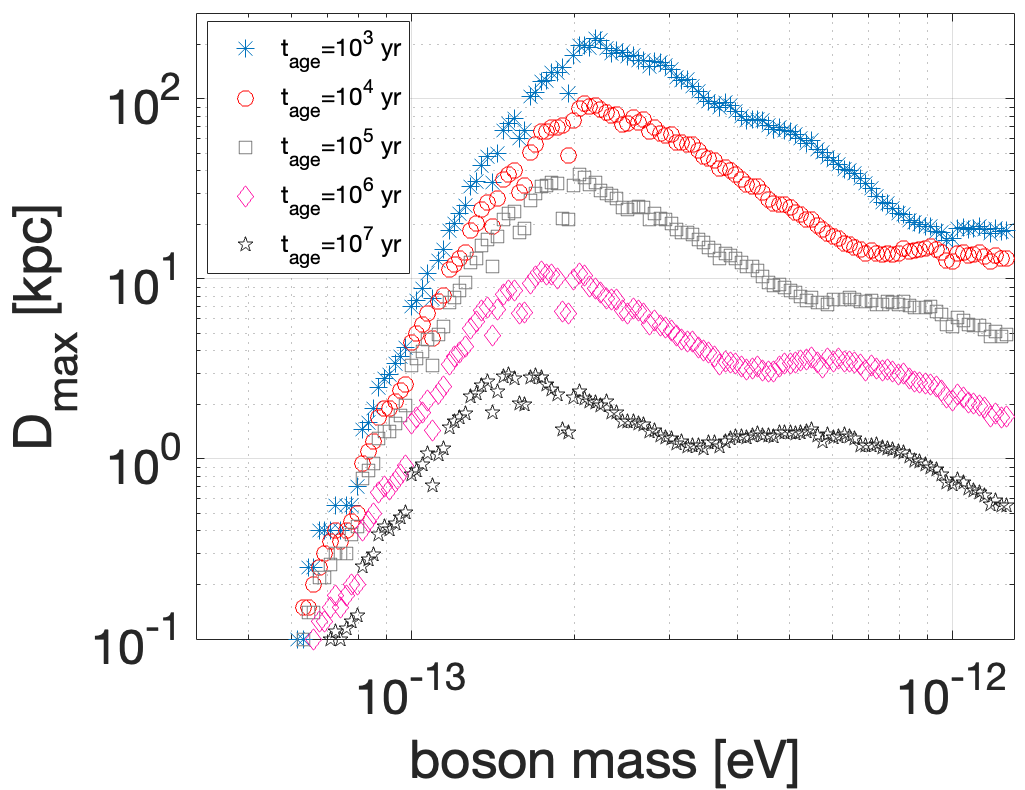}
\caption{Maximum %MDPI: Please change hyphen into minus sign -> Not possible
distance at which at least $5\%$ of a simulated population of BHs-boson cloud would produce a detectable GW signal in the search described in~\cite{LVKallskyBosonsO32021}. 
The simulated population has a maximum BH mass of $100 M_\odot$. The curve is produced for different markers for systems with ages between $10^3$ yrs and $10^7$ yrs, as indicated in the legend. Figure taken from~\cite{LVKallskyBosonsO32021}.}
\label{fig:bosondist}%MDPI: Please ensure that permission has been obtained and there is no copyright issue. If copyright is needed, please provide a citation in the following format: “Reprinted/adapted with permission from Ref. [XX]. Copyright year, copyright owner’s name”. More details on “Copyright and Licensing” are available via the following link: https://www.mdpi.com/ethics#10.
\end{figure}

The first all-sky survey for persistent, quasi-monochromatic GW signals emitted by ultralight scalar boson clouds around spinning BHs is reported in~\cite{LVKallskyBosonsO32021}. The search analyzed the frequency range $[20;610]~\rm Hz$ of the O3 observing run of Advanced LIGO. According to the expected spin wandering signal model from boson clouds, a small range around the $\dot{f}_{\rm gw}$ parameter has been considered. The main search has been carried out using the method in~\cite{DantonioMethBC2018}, and the most interesting candidates have been followed up with a FrequencyHough-based method and the Viterbi method. No potential CW candidate remains after the follow-up, and upper limits on the signal strain are provided at the $95\%$ confidence level.  The minimum value is $1.04 \times 10^{-25}$ in the most sensitive frequency region of the detectors ${\sim}130$ Hz. These upper limits on the strain can be translated into exclusion regions in the BH-boson cloud mass plane after assuming some parameters of the BH population: the age of the source, its distance and its spin.
Along with these exclusion regions, it is possible to derive the maximum distance at which a given BH-boson cloud system, with a certain age, is not emitting  CWs, as a function of the boson mass (see Figure \ref{fig:bosondist}). In other words, these maximum distances tell us how far we can exclude the presence of an emitting system according to the null detection results obtained by the search. 
These constraints have been computed by simulating a BH population with masses in the range $[5;100]~M_{\odot}$ and a uniform spin distribution in the range $[0.2;0.9]$. The maximum distance corresponds to the distance at which at least $5\%$ of the simulated signal produces a strain amplitude higher than the upper limits of the search, meaning that it would have been detected. The paper in~\cite{LVKallskyBosonsO32021} also reports the same figure but for a BH population with maximum masses of $50~M_{\odot}$. These constraints strongly depend on the ensemble properties of the simulated BH population. The plot in Figure \ref{fig:bosondist} shows that the presence of systems younger than $10^3$ yrs is disfavored within the Galaxy, if formed by bosons with masses above $1.2 \times 10^{-13}$ eV, for the considered BH mass distribution. Older systems, which produce a smaller strain amplitude, are more difficult to constrain.

The search in~\cite{LVKallskyBosonsO32021} is the first of this type, all-sky type and optimized for frequency wandering signals, although it is possible to compute the same exclusion regions of the BH/scalar boson mass parameter space also using upper limits from all-sky searches for spinning NSs as in~\cite{Palomba2019PRL}.
A previous directed search for boson clouds from Cygnus X-1 was performed in O2 data in~\cite{SunCygnus2020} using the same Markov model Viterbi method applied for the follow-up of the all-sky boson cloud O3 search~\cite{LVKallskyBosonsO32021}.

Another potential candidate for CW signals, in the subset of the ultralight DM particles with masses $10^{-14}$--$10^{-11}$ eV/c$^2$ is a vector boson, the dark photon, which directly couples to GW interferometers (see Section \ref{ssec:DMtheory}). The latest O3 LIGO-Virgo data all-sky search for the CW signature from this type of system provided in~\cite{LVK2021O3DPDM}.  The boson mass range probed by this search is   (2--4) $\times \,10^{-13}$ eV/c$^2$, corresponding to the detectors frequencies $[10;2000]~\rm Hz$.
The search applies two complementary methods.  One pipeline is based on a cross-correlation method~\cite{Pierce2018} and already used in a previous dark photon DM O1 search~\cite{Guo2019O1}. The second is a semi-coherent method~\cite{MillerDPmeth2021} based on the band-sampled-data framework~\cite{PiccinniBSD2018} and adapted from the one used in the latest O3 BH-boson cloud search~\cite{LVKallskyBosonsO32021,DantonioMethBC2018}.
No evidence of DM signatures has been found, and upper limits on the signal strain are derived. These limits can be converted into upper limits on the coupling factor of the interaction between the dark photon  and the baryons in the detector. These constraints surpass the ones provided by existing DM experiments, such as the Eöt--Wash torsion balance~\cite{Schlamminger2008Eot} and MICROSCOPE~\cite{Berge2018MICROSCOPE}, and improve previous O1 results by a \mbox{factor ${\sim}100$.}

Searches for CW signatures from a pair of CDOs or PBHs far from the coalescence are also possible. The only actual search for systems with a frequency and amplitude evolution described by Equations  (\ref{eq:fPBH}) and   (\ref{eq:h0PBH}) has been performed in O1 data~\cite{Horowitz2020CDO}. How these CDOs can be formed is still under debate. One option is to assume that these objects can be trapped by normal matter, for instance a planet or even the Sun. Once these objects are formed, binaries can form if orbiting close enough to each other.  CDOs emitting GWs from the inspiral phase inside a solar system object should be very dense and in a very close binary orbit. The search in~\cite{Horowitz2020CDO} searches for CW signals from CDOs orbiting near the center of the Sun in the GW frequency range 50--550 Hz. No signal is claimed, and upper limits are provided. These limits are converted into upper limits on the mass of CDOs.
For almost the full frequency range investigated, these limits are below $10^{-9}~M_{\odot}$ with a minimum value of $5.8 \times 10^{-10}~M_{\odot}$ at 525.5 Hz.

Constraints on the rates and abundances of nearby planetary- and asteroid-mass primordial BHs are provided in~\cite{LVKO3allsky2022,millerPBHO3aPowerflux2021}. These constraints are mapped from the upper limits strain results of the O3a Powerflux search and the O3 FrequencyHough search with minimal modeling assumptions on the PBH population. As anticipated in the previous paragraph, the search in~\cite{LVKO3allsky2022} targets CW from unknown NS, but the results can be generalized for the case of CDOs or PBHs far from the coalescence as far as they follow a linear frequency evolution. In principle, this search is able to probe GW signals below 250 Hz from inspiralling PBHs binaries with chirp masses smaller than $O(10^{-5})M_{\odot}$. The constraints obtained from the latest full O3 search in~\cite{LVKO3allsky2022} are more stringent than those presented in~\cite{millerPBHO3aPowerflux2021} from the O3a search thanks to the use of a longer observing time. Unfortunately, none of these results are stringent enough to be able to constrain the merging rate of PBHs but will certainly be interesting for future third-generation of detectors~\cite{PunturoET2010,reitze2019cosmic}.

\section{Conclusions}
\label{sec:future}
In  recent years, we have witnessed the huge impact that GW science had on our understanding of the Universe. Each detection has provided an interesting tool to test what already is known, but mostly the unknown, about these fascinating compact objects able to emit GWs. The main actors in these scenes have been the well-known merger events, either as a pair of BHs or a system involving at least one NS. %Please confirm intended meaning is retained. ->OK
 We have witnessed the birth of the so-called multi-messenger astronomy, an incredibly powerful way to depict the details of a binary NS merger through EM, neutrino and particle astronomy. 
However, this is just the beginning, and indeed, the increase in the detectors' sensitivity, expected for the upcoming fourth observing run, along with the improvement  of existing pipelines, will certainly bring about the discovery of even more fascinating and unexpected phenomena. 
In this scenario, CW signals can be the next surprise in GW astronomy.  A CW detection from NS could provide evidence for star deformations and shed light on the NS structure and their thermal, spin, and magnetic field evolution. On the other hand, if the CW detection happens to be in the case of a DM candidate such as from a boson cloud around spinning BH, this will lead to the first direct evidence of the existence of DM, solving one of the most intriguing open problems of astrophysics.

Unfortunately, to date, any search has been able to find strong evidence of GW emission from CW sources. However, an increasing number of known pulsars is surpassing their spin-down limit. The lowest constraints on the equatorial ellipticities are very close to realistic expectations for normal EOS. A comment is probably necessary to explain why, even using the most sensitive pipelines for CW searches, any GW has been observed from these sources. It is not unlikely that the assumed models are not well representative of the realistic emission scenario happening for CW, in particular for NS. This is the main reason why, along with improvements in the pipeline's sensitivity, many approaches tend to improve the pipeline's robustness with respect to the model. 
The final answer to this long-debated issue of the existence (or not) of CW signals is probably just around the corner, and the upcoming O4 run is expected from a wide community of scientists to finally solve this unanswered question. Even if no detection happens during O4, a new generation of detectors---Einstein Telescope, Cosmic Explorer, LISA and more~\cite{PunturoET2010,reitze2019cosmic,LISA,kalogera2021generation,kawamura2020decigo,Luo2016TianQin}---will start their observation in the future, providing even more interesting detection scenarios.

\vspace{6pt}

%%%%%%%%%%%%%%%%%%%%%%%%%%%%%%%%%%%%%%%%%%
%% optional
%\supplementary{The following are available online at \linksupplementary{s1}, Figure S1: title, Table S1: title, Video S1: title.}


%%%%%%%%%%%%%%%%%%%%%%%%%%%%%%%%%%%%%%%%%%

\funding{This work was supported by the L'Oréal-UNESCO fellowship for Women in Science.}

\acknowledgments{I thank Cristiano Palomba for the useful suggestions and discussions, improving the first version of this review. I also thank the Continuous Wave working group and the LIGO-Virgo-KAGRA Collaboration for the material used in this review.}

\conflictsofinterest{The author declares no conflict of interest.}

%%%%%%%%%%%%%%%%%%%%%%%%%%%%%%%%%%%%%%%%%%
%% Only for journal Encyclopedia
%\entrylink{The Link to this entry published on the encyclopedia platform.}

%%%%%%%%%%%%%%%%%%%%%%%%%%%%%%%%%%%%%%%%%%
%% Optional
%\abbreviations{The following abbreviations are used in this manuscript:\\

%\noindent 
%\begin{tabular}{@{}ll}
%BH & Black Hole \\
%CW & Continuous Wave\\
%NS & Neutron Star\\
%EOS & Equation Of State\\
%DM & Dark Matter\\
%PBH & Primordial Black Hole\\
%DP & Dark Photon\\
%GW & GWs\\
%EM & Electromagnetic\\
%LMXB & Low-Mass X-ray Binary\\
%O1 & Observing Run 1\\
%O2 & Observing Run 2\\
%O3 & Observing Run 3\\
%\end{tabular}
%}

%%%%%%%%%%%%%%%%%%%%%%%%%%%%%%%%%%%%%%%%%%
%% Optional
\appendixtitles{no} % Leave argument ``no'' if all appendix headings stay EMPTY (then no dot is printed after ``Appendix A''). If the appendix sections contain a heading then change the argument to "yes".
\appendixstart
\appendix

\begin{adjustwidth}{-\extralength}{0cm}
\printendnotes[custom] 
%\centering %% If there is a figure in wide page, please release command \centering
\reftitle{References}

\end{adjustwidth}

\end{document}